\documentclass[useAMS,usenatbib]{mn2e}
\usepackage{amsmath,amssymb,txfonts,graphicx,color,enumitem,soul,enumitem,xfrac,mathtools,tikz,textcomp,bm,ifthen,todonotes,units,etoolbox,eso-pic}
\usepackage{journals,macros}
\usepackage[pass,letterpaper]{geometry}
\usepackage[euler]{textgreek}
\pdfoutput=1
\setcitestyle{comma}

\usepackage{soul}

\newboolean{shownotes}
\setboolean{shownotes}{true}

\newcommand\esnote[1]{%
\ifthenelse{\boolean{shownotes}}
{%
\note{green!20}{Suchyta}{#1}
}
{}}

\newcommand\emhnote[1]{%
\ifthenelse{\boolean{shownotes}}
{%
\note{orange!20}{Huff}{#1}
}
{}}

\newcommand\janote[1]{%
\ifthenelse{\boolean{shownotes}}
{%
\note{yellow!20}{Aleksi\'c}{#1}
}
{}}

\usepackage[colorlinks,linkcolor=blue,citecolor=blue,urlcolor=blue,pdftitle={Balrog},pdfauthor={Suchyta et al.},pdfproducer={}]{hyperref}

\begin{document}

\date{}


\title{No galaxy left behind: accurate measurements with the faintest objects in the Dark Energy Survey}



\author[Suchyta, Huff, Aleksi\'c et al. (DES Collaboration)]{
\parbox{\textwidth}{
\Large
E.~Suchyta$^{1,2\star}$,
E.~M.~Huff$^{1,2\dagger}$,
J.~Aleksi\'c$^{3}$,
P.~Melchior$^{1,2}$,
S.~Jouvel$^{4}$,
N.~MacCrann$^{5}$,
A.~J.~Ross$^{2}$,
M.~Crocce$^{6}$,
E.~Gaztanaga$^{6}$,
K.~Honscheid$^{1,2}$,
B.~Leistedt$^{4}$,
H.V.~Peiris$^{4}$,
E.~S.~Rykoff$^{7,8}$,
E.~Sheldon$^{9}$,
T.~Abbott$^{10}$,
F.~B.~Abdalla$^{4,11}$,
S.~Allam$^{12}$,
M.~Banerji$^{13,14}$,
A.~Benoit-L{\'e}vy$^{4}$,
E.~Bertin$^{15,16}$,
D.~Brooks$^{4}$,
D.~L.~Burke$^{7,8}$,
A.~Carnero~Rosell$^{17,18}$,
M.~Carrasco~Kind$^{19,20}$,
J.~Carretero$^{3,6}$,
C.~E.~Cunha$^{7}$,
C.~B.~D'Andrea$^{21}$,
L.~N.~da Costa$^{17,18}$,
D.~L.~DePoy$^{22}$,
S.~Desai$^{23,24}$,
H.~T.~Diehl$^{12}$,
J.~P.~Dietrich$^{23,25}$,
P.~Doel$^{4}$,
T.~F.~Eifler$^{26,27}$,
J.~Estrada$^{11}$,
A.~E.~Evrard$^{28,29}$,
B.~Flaugher$^{12}$,
P.~Fosalba$^{6}$,
J.~Frieman$^{12,30}$,
D.~W.~Gerdes$^{29}$,
D.~Gruen$^{25,31}$,
R.~A.~Gruendl$^{19,20}$,
D.~J.~James$^{10}$,
M.~Jarvis$^{26}$,
K.~Kuehn$^{32}$,
N.~Kuropatkin$^{12}$,
O.~Lahav$^{4}$,
M.~Lima$^{17,33}$,
M.~A.~G.~Maia$^{17,18}$,
M.~March$^{26}$,
J.~L.~Marshall$^{22}$,
C.~J.~Miller$^{28,29}$,
R.~Miquel$^{3,34}$,
E.~Neilsen$^{12}$,
R.~C.~Nichol$^{21}$,
B.~Nord$^{12}$,
R.~Ogando$^{17,18}$,
W.J.~Percival$^{21}$,
K.~Reil$^{8}$,
A.~Roodman$^{7,8}$,
M.~Sako$^{26}$,
E.~Sanchez$^{35}$,
V.~Scarpine$^{12}$,
I.~Sevilla-Noarbe$^{19,35}$,
R.~C.~Smith$^{10}$,
M.~Soares-Santos$^{12}$,
F.~Sobreira$^{12,17}$,
M.~E.~C.~Swanson$^{20}$,
G.~Tarle$^{29}$,
J.~Thaler$^{36}$,
D.~Thomas$^{21}$,
V.~Vikram$^{37}$,
A.~R.~Walker$^{10}$,
R.~H.~Wechsler$^{7,8,38}$,
Y.~Zhang$^{29}$
\begin{center} (The DES Collaboration) \end{center}
}
\vspace{0.4cm}
\\
\parbox{\textwidth}{\rm
{\Large $^{\star}$} \href{mailto:eric.d.suchyta@gmail.com}{eric.d.suchyta@gmail.com} \\
{\Large $^{\dagger}$} \href{mailto:emhuff@gmail.com}{emhuff@gmail.com} \\
\hyperlink{affil}{\color{black} Author affiliations are listed at the end of this paper.}
}
}

\maketitle

\begin{abstract}

  Accurate statistical measurement with large imaging surveys has
  traditionally required throwing away a sizable fraction of the
  data. This is because most measurements have relied on
  selecting nearly complete samples, where variations in the
  composition of the galaxy population with seeing, depth, or other
  survey characteristics are small.
  
  We introduce a new measurement method that aims to minimize this
  wastage, allowing precision measurement for any class of detectable stars or
  galaxies. We have implemented our
  proposal in \balrog{}, software which embeds fake objects
  in real imaging to accurately characterize measurement
  biases.

  We demonstrate this technique with an angular clustering measurement
  using Dark Energy Survey (DES) data. We first show that recovery of
  our injected galaxies depends on a variety of survey
  characteristics in the same way as the real data. We then construct
  a flux-limited sample of the faintest galaxies in DES, chosen
  specifically for their sensitivity to depth and seeing
  variations. Using the synthetic galaxies as randoms in the 
  Landy-Szalay estimator suppresses the effects
  of variable survey selection by at least two orders of
  magnitude. With this correction, our measured angular clustering is
  found to be in excellent agreement with that of a matched sample
  from much deeper, higher-resolution space-based Cosmological Evolution Survey (COSMOS)
  imaging; over angular scales of $0.004\degr < \theta < 0.2\degr$,
  we find a best-fit scaling amplitude between the DES and COSMOS
  measurements of $1.00 \pm 0.09$.

  We expect this methodology to be broadly useful for extending measurements'
  statistical reach in a variety of upcoming
  imaging surveys.
\end{abstract}
\nokeywords

\section{Introduction}

Wide-field optical surveys have played a central role in modern astronomy. 
The Sloan Digital Sky Survey \citep[SDSS,][]{sdss-overview} alone has furnished
nearly 6,000~publications across a wide variety of subjects: from star formation,
to galaxy evolution, to measuring cosmological parameters; among a multitude of others.
The discovery of cosmic acceleration
\citep{Riess_Snae, Perlmutter_Snae} has motivated several expansive
imaging surveys for the future: for instance, the Large Synoptic Survey
Telescope,\footnote{\href{http://www.lsst.org/lsst/}{http://www.lsst.org/lsst/}}
the Wide-Field Infrared Survey Telescope \citep{WFIRST}, and Euclid
\citep{Euclid}.
The legacy of these next-generation imaging efforts will almost certainly yield an
even richer harvest than what has come before them.


With large surveys, astronomical 
sample sizes have grown, increasing the statistical power
of their measurements;
\textit{with great power comes great responsibility}%
\footnote{Though we have referenced \citet{spiderman} as an example, we note, 
the phrase did not originate with Spiderman.
The quote is often attributed to different sources, including (likely incorrectly) Voltaire,
and can be traced back as far as at least the Gospel of Luke (12:48).}
\citep[see e.g.][]{spiderman} 
for control of systematic errors. Taking full advantage of these data means ensuring
that the precision of these measurements is matched by their
accuracy. At present time, however, high-precision
measurements are generally made with samples drawn from only the
fraction of the data that is nearly complete. We argue that the
current state of the art in survey astronomy is in many ways wasteful
of information, and lay out a general method for improvement.

This paper focuses on measurements of the galaxy angular correlation
function for highly incomplete, flux-limited samples of galaxies,
especially near the detection threshold.  We have chosen this approach for two reasons.
First, this measurement is an especially challenging example of
systematic error mitigation; we show below that, for our faintest
galaxies, we will have to eliminate systematic biases that are much
larger than our signal, and do so over a wide range of survey
conditions. The second reason is that systematic effects
relevant for angular clustering  measurements also directly impact probes
of cosmic acceleration \citep{Weinberg_Review}, where the requirements
on systematic error control are particularly strict.

\subsection{The current state of the art}
Astronomers have been measuring galaxy clustering for several decades,
since at least \citet{super-old-zwicky}.
The angular two-point correlation function, $w(\theta)$, is
a common tool used to characterize the anisotropies in the galaxy ensemble. 
From the very beginning, efforts to measure $w(\theta)$ have been challenged by the presence of
anisotropies in the data arising from imperfect measurements, or from
astrophysical complications unrelated to large-scale structure.

A complete list of sources of systematic effects is difficult (if not impossible) to
compile, but some issues are
common to all extragalactic measurements, like \sg{} separation and
photometric calibration.  
Because the point spread function (PSF) varies across the
survey area, the accuracy with which
galaxies can be distinguished from stars will vary,
introducing anisotropies associated with stellar contamination.
Accurate, uniform photometric calibration for a multi-epoch wide-field
optical survey is difficult to accomplish
\citep{PS1_photometric_calibration}, and given the variations in
seeing, airmass, transparency, and other observing conditions,
uniform depth is generally unachievable. A wide variety of
schemes have been used to ameliorate these complicating effects.

For a $w(\theta)$ measurement with the Automated Plate Measurement
survey -- among the earliest digitized sky surveys --
\citet{old_school} built models of the selection function, including
plate measurement effects (e.g., the variation of the photographic
emulsion's sensitivity across each plate), observational effects
(atmospheric extinction) and astrophysical effects (Galactic
extinction). For each of these, they estimated the contribution of the
systematic effect to the final $w(\theta)$ measurement. Stellar
contamination was dealt with by subtracting estimated stellar
densities from the map of galaxy counts in cells, and adjusting the
amplitude of the final $w(\theta)$ measurement to compensate for the
estimated dilution due to stellar contamination.

Similar measurements of $w(\theta)$ were made for validation purposes
in the early SDSS data
\citep{SDSS_EDR}. The authors here cross-correlated the measured
galaxy densities with a number of known sources of systematic errors
in order to determine which regions of the survey to mask.

Many subsequent SDSS analyses were based on a volume-limited sample of
luminous red galaxies, from which $\sim\!120,000$ objects were
targeted for SDSS spectroscopy \citep{LRG_sample}. Here again (see
also \citealt{lrg_photo_clustering} for the properties of the parent
photometric sample) the strategy was to use cross-correlation
techniques to remove data that would imperil the analysis, leaving an
essentially complete sample.

The targets selected for the larger SDSS-III Baryon Oscillation
Spectroscopic Survey (BOSS) measurements \citep{BOSS} were
substantially fainter, and the systematic error corrections for these
samples necessarily more sophisticated.  \citet{ameliorate} explored
several mitigation strategies for SDSS data.  A linear model for the
dependence of the galaxy counts as a function of potential sources of
systematic errors was built, allowing for subtraction of the
systematic effects from the final galaxy $w(\theta)$ measurement.  For
the most important systematic effects (constrained again by
cross-correlation with the galaxies), galaxies in the $w(\theta)$
estimator were upweighted by the inverse of their detection
probability.  The BOSS baryon acoustic oscillation scale measurement
in \citet{BOSS_BAO} made use of this weighting scheme. With the
exception of stellar occultation, these effects were mostly
perturbative, and the errors on the angular clustering were large
enough that the stellar occultation corrections only had to be
characterized at the $\sim\!10\%$ level.

The imaging systematic error mitigation used by the WiggleZ
spectroscopic survey \citep{blake2010} came closest to the spirit of
this paper. Their spectroscopic target catalog was built by a
combination of SDSS and Galaxy Evolution Explorer%
\footnote{\href{http://www.galex.caltech.edu/}{http://www.galex.caltech.edu/}}
(GALEX) measurements. The blue emission-line galaxies targeted by
WiggleZ were faint enough to be substantially affected by variations
in the SDSS completeness, so the GALEX catalogs were used to estimate
the variation of the target selection probability with various survey
properties. Models were fit to this dependence, and the results were
directly incorporated into the window function used in power spectrum
estimation. The resulting corrections had a $\sim\!0.5\sigma$ effect
on the final power spectrum, and so like SDSS only needed to be
accurate at the $\sim\!10\%$ level.

This list is not exhaustive, but we believe it gives a fair picture of
the state of the art. Generally, for their extragalactic clustering
measurements, modern photometric surveys have relied on selecting a
relatively complete sample, and then applying small corrections late
in the analysis. We believe that this approach is a poor fit to the
age of precision cosmology with `big data.' The rest of this paper
will present our proposed alternative.

\subsection{Modeling the Dark Energy Survey selection function}

We propose to measure the selection function of imaging surveys by
embedding a realistic ensemble of fake star and galaxy images in the
real survey data.  The resulting measurement catalogs comprise a Monte Carlo
sampling of the selection function and measurement biases of the
survey, and can naturally account for systematic effects arising from the
photometric pipeline, detector defects, seeing, and other sources of
observational systematic errors.  Several of the major systematic
errors examined in the above measurements can be straightforwardly
estimated and removed using the embedded catalogs, though
astrophysical effects like dust and photometric calibration must of
course be modeled using external data.

We test this technique using Dark Energy Survey (DES) imaging. DES is
a $5-$year optical and near-infrared survey of $5,000~\mathrm{deg}^2$
of the South Galactic Cap, to $i_{AB}\le 24$ \citep{des2005}.  The
survey instrument, the Dark Energy Camera \citep[DECam,][]{decam}, was
commissioned in fall 2012.  During the Science Verification (SV)
phase, which lasted from November 2012 to February 2013, data were
taken over $\sim\!250~{\rm deg}^2$ in a manner mimicking the full
5-year survey, but with substantial depth variations (see
e.g. \citealt{boris_maps}), mainly due to weather and early DECam
operational challenges.  Coadd images in each of the five bands, as
well as a detection image combining the $riz$ filters, were produced
from the $\sim\!10$ single-epoch exposures per filter.

Our work is complementary to that of \citet{ufig-des}, who used
generative modeling, in combination with outputs from the Blind
Cosmology Challenge \citep{bcc} and the Ultra Fast Image Generator
\citep{ufig}, to simulate DES-like data which were then run through the
DES analysis pipeline \citep{mohr2012, desai2012}. A fully generative
approach does have some advantages over the Monte Carlo sampling of
the images described here. With a generative model, one can explore
counterfactual realizations of the survey. This helps, for instance,
in mapping out the interaction between the survey selection function
and the galaxy population (for instance, how the angular clustering of
galaxies interacts with the deblending and sky-subtraction
algorithms).  By construction, our embedding strategy considers only
the single DES-realization of the survey properties.

However, the generative modeling approach is more sensitive 
to model mis-specification errors; it requires models not
only for the noise, photometric calibration, star and galaxy ensemble
properties, etc., but also for cosmic rays, bright stellar diffraction
spikes, CCD defects, satellite trails, and other non-physical
signatures that are difficult to model accurately. The embedded
simulations, by contrast, inherit many of the properties of the image
that are otherwise difficult to model. To keep the embedded population
as realistic as possible, we draw our simulated stars and galaxies
from catalogs made from high-resolution Hubble Space Telescope
imaging.

\subsection{Angular clustering in the Dark Energy Survey}

\citet{benchmark} present a DES \textit{benchmark} measurement of
$w(\theta)$, adopting a standard approach to their clustering analysis
by choosing a relatively complete sample ($i < 22.5$) and masking
potential sources of systematic errors traced by maps of the DES
observing properties measured by \citet{boris_maps}.  In this paper,
we use our Monte Carlo simulation framework to correct for the
spatially-dependent completeness inhomogeneities, and then measure
clustering signals at magnitudes well below the nominal limiting depth
of $i < 22.5$ used by \citet{benchmark}.

The paper is organized as follows.  In \autoref{sec:implementation},
we present \balrog{},%
\footnote{\href{https://github.com/emhuff/Balrog}{https://github.com/emhuff/Balrog}.
  \balrog{} is \textbf{\textit{not}} an acronym.  The software was
  born out of the authors \textit{digging too deeply and too greedily}
  into their data, ergo the name.}
our software pipeline for embedding simulations into astronomical
images.  In \autoref{sec:sample}, we describe our empirical procedure
for generating a realistic ensemble of simulated sources, then
prototype \balrog{} by injecting $\sim\!40,000,000$ simulated objects
into $178~\mathrm{deg}^2$ of DES SV coadd images.  We generate a
synthetic catalog using the same procedure as is used for generation
of the DES science catalogs.  \autoref{sec:validation} validates that
the photometric properties of the synthetic catalogs are a close match
to those of the real DES catalogs for a wide range of quantities.  If
these synthetic catalogs really capture the variation in the survey
selection function and measurement biases, it should be possible to
use them as randoms to measure $w(\theta)$ accurately even for the
faintest galaxies in the survey. We do exactly this in
\autoref{sec:measurements}, demonstrating that our clustering
measurements for the faintest DES galaxies ($23< i < 24$) show
excellent agreement with higher resolution external space-based data,
which are complete over the selection range.  
The shapes of our $w(\theta)$ curves match general expectation.
\autoref{sec:discussion} concludes with a discussion of our results.

\section{\balrog{} Implementation}
\label{sec:implementation}

\usetikzlibrary{shapes, arrows, positioning, calc, decorations.pathreplacing}
\tikzset{%
  input/.style = {draw,fill=gray, fill opacity=0.0, text opacity=1, trapezium,trapezium left angle=70,trapezium right angle=-70, font=\small},
  output/.style = {draw,fill=gray, fill opacity=0.6, text opacity=1, trapezium,trapezium left angle=70,trapezium right angle=-70, font=\small},
  com/.style = {draw, fill=gray, fill opacity=0.3, text opacity=1, thick, rectangle, font=\small},
  dec/.style = {draw, fill=gray, fill opacity=0.3, text opacity=1, diamond, font=\small},
  gs/.style = {draw, fill=gray, fill opacity=0.3, text opacity=1, thick, rectangle, font=\small},
  sex/.style = {draw, fill=gray, fill opacity=0.3, text opacity=1, thick, rectangle, font=\small}
}

\begin{figure*}
\centering
\begin{tikzpicture}[auto, thick, >=triangle 45]

\draw node [input, align=center, text width=2.8cm] (mconf) {\textbf{Measurement software configurations}};
\draw node [sex, right=0.5cm of mconf, align=center, text width=3.4cm] (rm2) {\textbf{Run measurement software}};
\draw node [output, above=1cm of rm2, align=center, text width=2.3cm] (truth) {\textbf{Simulated objects truth catalog}};
\draw node [com, right=3.8cm of truth, align=center, text width=2.5cm] (parse) {\textbf{Parse simulation configuration}};
\draw node [input, above=0.5cm of parse, align=center, text width=2.5cm] (pyconfig) {\textbf{Object simulation configuration file}};
\draw node [input, above=2.8cm of mconf, align=center, text width=3.5cm] (imaging) {\textbf{Data: image, weight map, PSF, zeropoint, gain}};
\draw node [output, right=0.5cm of rm2, align=center, text width=1.8cm] (simimage) {\textbf{Data $+$ simulations}};
\draw node [sex, above=1cm of mconf, align=center, text width=3.5cm] (rm1) {\textbf{Run measurement software}};
\draw node [output, below=0.7cm of rm2, align=center, text width=3cm] (sim) {\textbf{Measurement catalog with simulations}};
\draw node [gs, right=0.4cm of simimage, align=center, text width=2.9cm] (draw) {\textbf{Draw simulated objects into image}};
\draw node [output, left=2cm of sim, align=center, text width=3cm] (nosim) {\textbf{Measurement catalog without simulations}};

\draw[->](pyconfig) -- node {} (parse);
\draw[->](parse) -- node {} (draw);
\draw[->](parse) -- node {} (truth);
\draw[->](mconf) -- node {} (rm1);
\draw[->](mconf) -- node {} (rm2);
\draw[->](truth) -- node {} (rm1);
\draw[->](truth) -- node {} (rm2);
\draw[->](imaging) -- ($(imaging.east) + (10.5,0)$) |- node {} (draw);
\draw[->](draw) -- node {} (simimage);
\draw[->](simimage) -- node {} (rm2);
\draw[->](imaging) -- node {} (rm1);
\draw[->](rm1) -- ($(rm1.west) - (2.0,0)$) |- node {} (nosim);
\draw[->](rm2) -- node {} (sim);

\end{tikzpicture}
\caption{High-level overview of \balrog{}'s processing. 
Shape usage follows standard flowchart notation.
White parallelograms are inputs, dark gray parallelograms are outputs, and light gray rectangles are processes/commands.
(The simulation truth catalog is coupled with the measurement software because by default \balrog{} runs \sex{} in association mode, 
using the simulation positions as the matching list,
cf. \autoref{sec:measurement}.)
}
\label{fig:flowoverview}
\end{figure*}
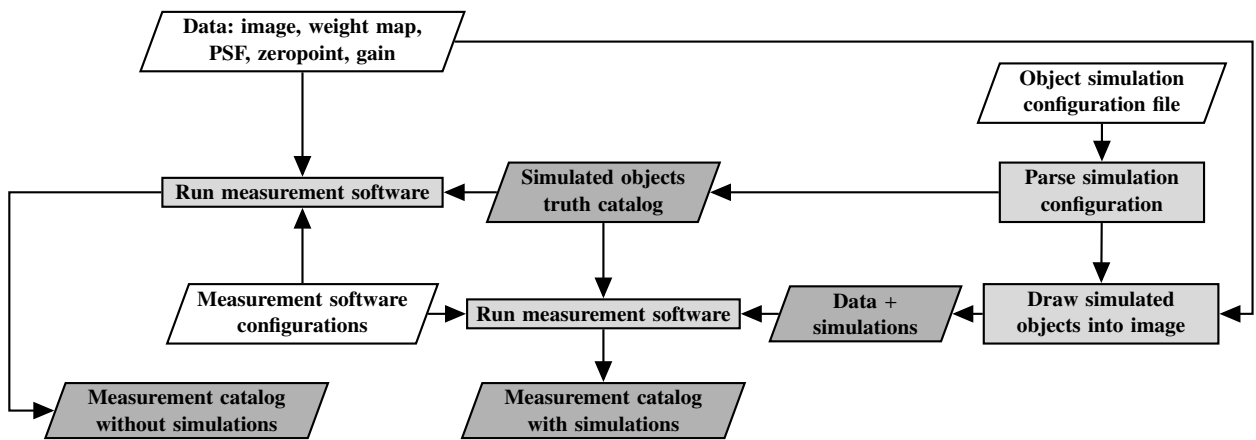

\balrog{} is a Python-based software package for embedding simulations
into astronomical images; \autoref{fig:flowoverview} shows a diagram
of the pipeline's workflow.  \balrog{} begins with an observed survey
image, then inserts simulated objects with known truth properties into
the image.  Source detection and analysis software is run over the
image, measuring the observed properties of the simulated objects.  We
emphasize that because a real survey image has been used, \balrog{}'s
output catalog automatically inherits otherwise difficult to simulate
features, such as over-subtraction of the sky background by the
measurement software, proximity effects of nearby objects, unmasked
cosmic rays, etc.

The remainder of this section further details how we implement these
injection simulations in \balrog{}.  The discussion is organized
according to three components of \balrog{}'s functionality, each of
which is devoted a section to follow:
\begin{enumerate}
\item input survey information, such as reduced images, their PSFs, and flux calibrations (\autoref{sec:surveyinfo});
\item simulation specifications, defining how to generate the simulated object population (\autoref{sec:simulation});
\item measurement software (\autoref{sec:measurement}).
\end{enumerate}

We have designed \balrog{} with ease of use and generality in mind,
allowing for a wide range of simulation implementations, and we provide thorough
documentation with the software.  \balrog{} employs software widely
used throughout the astronomical community: internally it calls \sextractor{}
\citep{sextractor} for source detection and measurement, and the object
simulation framework is built on \galsim{}
\citep{galsim}.

\subsection{Survey information}
\label{sec:surveyinfo}

The top left of \autoref{fig:flowoverview} lists the survey data
required by \balrog{}.  First are the reduced images and their 
weight maps -- the inverse of the noise variance of the image at background level.  
The latter are required for reliable measurements of object properties; 
\balrog{} does not modify the weight maps, but passes them
as input arguments to \sex{}.  Both the images and weight maps are
expected to conform to the Flexible Image Transport System (FITS)
standard \citep{fits, fitswcs}.

All simulated \balrog{} objects are convolved with a PSF prior to
being drawn into the image.  Currently, \balrog{} requires a PSF model
generated by \psfex{} \citep{psfex} to be given as the input defining
the convolution kernel.  These models encode a set of basis images to
represent the spatial-dependence of the PSF, with an adjustable-degree
polynomial for interpolation of the basis coefficients across the
image.  \balrog{}'s PSF convolution calls \galsim{}'s
\texttt{Convolve} method, and the implementation operates in World
Coordinates, where the astrometric solution to use is read from the
image's FITS header.  We note that \galsim{}'s PSF functionality is
not limited to images generated by \psfex{}; it accepts a wide variety
of other possibilities as well.  We have chosen to implement the
\psfex{} models in our initial version of \balrog{}, because they are
used in DES.  However, \balrog{} could be extended to accept a broader
range of PSF model types.

A photometric zeropoint ($z_{p}$) is required to transform simulated
object magnitudes ($m$) into image fluxes ($F$), by applying the usual
conversion between the two quantities:
\begin{equation}
F = 10^{\scalebox{0.95}{ ($z_p - m) / 2.5$ }}.
\end{equation}
Natively, the conversion assumes that all pixels share this same
calibration,\footnote{With \balrog{}'s user-defined function API, one
  can implement non-uniform photometric calibrations across an image,
  such as we do in \autoref{sec:desbalrog} with stellar locus
  regression zeropoint offsets.  We refer readers to the code
  repository and documentation therein for details.}  whereby the
images should have standard reductions, such as bias subtraction and
flat field division, applied prior to running \balrog{} (in order to
remove pixel-dependent variations across the image).  By default,
\balrog{} tries to read the zeropoint from the FITS header, but also
accepts command line arguments.

In addition to the noise inherited from the image, \balrog{} also adds
Poisson noise to the simulated objects' pixel flux values, where the
noise level is set by the image's effective electron/ADU gain.  This
added Poisson noise is only significant when the object flux level is
well above the background variation level.  Like the zeropoint,
\balrog{} can read the gain from the FITS header or accept a command
line argument.

\subsection{Simulating images}
\label{sec:simulation}


\usetikzlibrary{shapes, arrows, positioning, calc, decorations.pathreplacing}
\tikzset{%
  input/.style = {draw,fill=white, fill opacity=0.25, text=black, text opacity=1, trapezium,trapezium left angle=70,trapezium right angle=-70, font=\small},
  output/.style = {draw,fill=gray, fill opacity=0.6, text=black, text opacity=1, trapezium,trapezium left angle=70,trapezium right angle=-70, font=\small},
  com/.style = {draw, fill={rgb:red,4;green,2;yellow,1}, fill opacity=0.1, text=black, text opacity=1, thick, rectangle, font=\small},
  dec/.style = {draw, fill={rgb:red,4;green,2;yellow,1}, fill opacity=0.1, text=black, text opacity=1, diamond, font=\small},
  gs/.style = {draw, fill={rgb:red,4;green,2;yellow,1}, fill opacity=0.4, text=black, text opacity=1, thick, rectangle, font=\small},
  sex/.style = {draw, fill=pink, fill opacity=0.7, text=black, text opacity=1, thick, rectangle, font=\small}
}

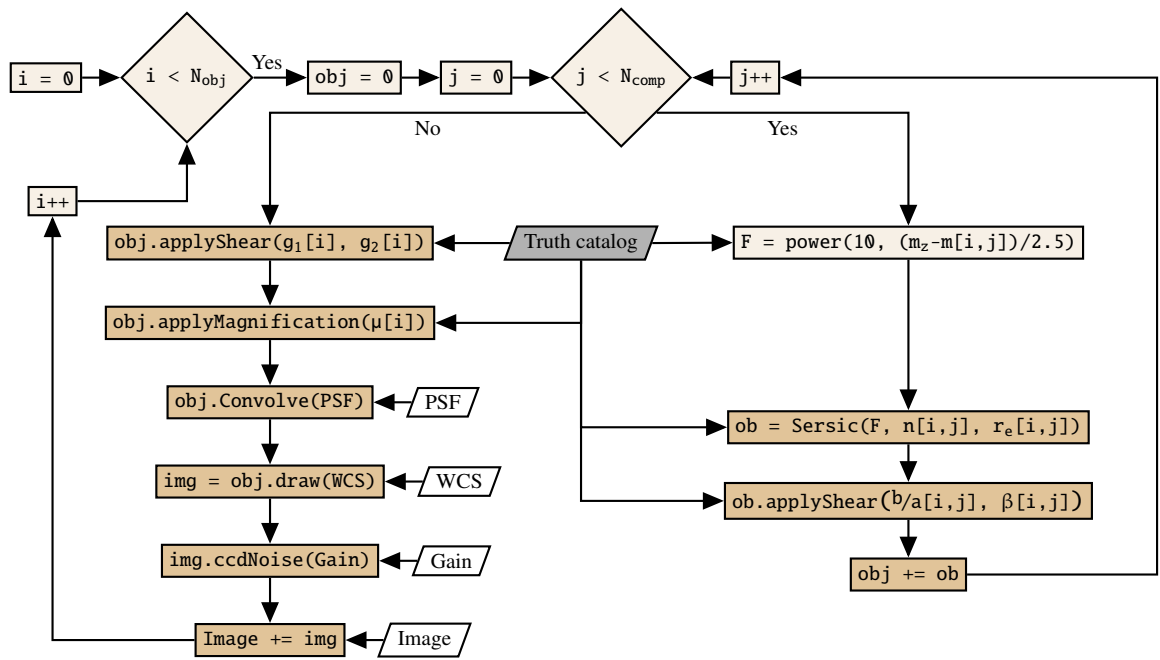
\begin{figure*}
\centering
\begin{tikzpicture}[auto, thick, >=triangle 45]

\draw

	node [com] (i0) {\texttt{i = 0}}
	node [dec, right=0.5cm of i0] (idec) {\texttt{i < $\texttt{N}_{\texttt{obj}}$}}
	node [com, right=0.7cm of idec] (obj0) {\texttt{obj = 0}}
	node [com, right=0.5cm of obj0] (j0) {\texttt{j = 0}}
    node [dec, right=0.5cm of j0] (jdec) {\texttt{j < $\texttt{N}_{\texttt{comp}}$}}
    
    node [gs, below left=1.5cm and 2.0cm of jdec] (shear) {\texttt{obj.applyShear($\texttt{g}_{\texttt{1}}$[i], $\texttt{g}_{\texttt{2}}$[i])}}
    node [gs, below=0.6cm of shear] (mag) {\texttt{obj.applyMagnification(\texttt{\textmu}[i])}}
    node [gs, below=0.6cm of mag] (convolve) {\texttt{obj.Convolve(PSF)}}
    node [gs, below=0.6cm of convolve] (draw) {\texttt{img = obj.draw(WCS)}}
    node [gs, below=0.6cm of draw] (noise) {\texttt{img.ccdNoise(Gain)}}
    node [gs, below=0.6cm of noise] (img+) {\texttt{Image += img}}
    node [com, below left=1cm and 1cm of idec] (i++) {\texttt{i++}}
    
    node [output, right=1.0cm of shear] (truth) {Truth catalog}
    node [input, right=0.5cm of convolve] (psf) {PSF}
    node [input, right=0.5cm of draw] (wcs) {WCS}
    node [input, right=0.5cm of noise] (gain) {Gain}
    node [input, right=0.5cm of img+] (image) {Image}
    
    node [com, below right=1.5cm and 1.0cm of jdec] (flux) {\texttt{F = power(10, ($\texttt{m}_{\texttt{z}}$-m[i,j])/2.5)}}
    node [gs, below=2.0cm of flux, align=center] (sersic) {\texttt{ob = Sersic(F, n[i,j], $\texttt{r}_{\texttt{e}}$[i,j])}}
    node [gs, below=0.5cm of sersic] (ellip) {\texttt{ob.applyShear{\large (}\nicefrac{\footnotesize \texttt{b}}{\footnotesize \texttt{a}}[i,j], \textbeta[i,j]{\large )}}}
    node [gs, below=0.5cm of ellip] (obj+) {\texttt{obj += ob}}
    node [com, right=0.5 of jdec] (j++) {\texttt{j++}}
    ;
    
	\draw[->](i0) -- node {}(idec);
 	\draw[->](idec) -- node[above,near start] {\footnotesize Yes} (obj0);
	\draw[->](obj0) -- node {} (j0);
	\draw[->](j0) -- node {} (jdec);
	
	\draw[->](jdec.south west) -| node[below,near start] {\footnotesize No} (shear.north);
	\draw[->](shear) -- node {} (mag);
	\draw[->](mag) -- node {} (convolve);
	\draw[->](convolve) -- node {} (draw);
	\draw[->](draw) -- node {} (noise);
	\draw[->](noise) -- node {} (img+);
	\draw[->](img+) -| node {} (i++);
	\draw[->](i++) -| node {} (idec.south);
	
	\draw[->](jdec.south east) -| node[below,near start] {\footnotesize Yes} (flux.north);
	\draw[->](flux) -- node {} (sersic);
	\draw[->](sersic) -- node {} (ellip);
	\draw[->](ellip) -- node {} (obj+);
	\draw[->](obj+.east) -- ($(obj+.east) + (2.5,0)$) |- node{} (j++.east);
	\draw[->](j++) -- node {} (jdec.east);
	
	\draw[->](truth) -- node{} (shear.east);
	\draw[->](truth.south) |- node{} (mag.east);
	\draw[->](truth.east) -- node{} (flux.west);
	\draw[->](truth.south) |- node{} (sersic.west);
	\draw[->](truth.south) |- node{} (ellip.west);
	\draw[->](psf) -- node{} (convolve);
	\draw[->](wcs) -- node{} (draw);
	\draw[->](gain) -- node{} (noise);
	\draw[->](image) -- node{} (img+);

\end{tikzpicture}
\caption{\balrog{}'s object simulation schema.
This figure is effectively a ``zoom in'' of the ``Draw simulated objects into image'' node of 
\autoref{fig:flowoverview}.
The truth catalog is generated by \balrog{} based on the user's configuration setup.
White parallelograms are inputs to the pipeline,
dark orange rectangles call \galsim{} commands,
and light orange nodes are \py{} code.
Diamonds are decision points.
There are two loops:
index \texttt{i} loops over the number of simulated objects, $\texttt{N}_\texttt{obj}$;
index \texttt{j} loops over the number of \sersic{} components for each object, $\texttt{N}_\texttt{comp}$.
The final output is the image in the bottom left of the diagram, after all the simulated objects have been embedded.
}
\label{fig:flowsim}
\end{figure*}

The right side of \autoref{fig:flowoverview} depicts image simulation
and injection.  \hypertarget{hyper:truth}{} \balrog{} simulates
objects as a superposition of arbitrarily many elliptical \sersic{}
profiles.  Users are free to assign the magnitude, half light radius,
\sersic{} index, orientation angle and axis ratio of each \sersic{}
component. (To be explicitly clear, the \sersic{} quantities are pre-convolution values.)
Each object also includes three adjustable quantities that
are shared between the components: a center coordinate, lensing shear,
and magnification.

\hypertarget{hyper:parse}{} Assigning object properties is
accomplished by Python code inside a configuration file which
\balrog{} parses and executes.  We have packaged example configuration
files with the software to demonstrate its usage: for instance,
assigning to constants, arrays, or jointly sampling from a
catalog. Users are also able to write any Python function of their own
and use it as a sampling rule, allowing generality and arbitrary
complexity to the simulations.

\hypertarget{hyper:draw}{} \balrog{} uses \galsim{} to perform all
the routines necessary to transform a catalog of truth quantities into
images of these simulated objects.  \galsim{} rendering is extensively
validated in \citet{galsim}, and demonstrated to be accurate enough
for simulation of weak lensing data in Stage III and IV dark energy
surveys, including DES.  Beyond accuracy alone, \galsim{} is ideal for
\balrog{} because it is highly modular; \balrog{}'s range of
simulation customizations are built upon this modularity.

Here, we overview the most important simulation steps in \balrog{},
and refer readers to the \balrog{} code repository and \galsim{}
documentation for complete details.  \autoref{fig:flowsim} is a
diagram summarizing the process.  In the text, our convention is to
denote \texttt{GalSim methods} using \texttt{typewriter font}.  First,
each \sersic{} component is initialized as a circularly symmetric
\texttt{Sersic} object, with a given flux, half light radius, and
\sersic{} index (right side of \autoref{fig:flowsim}).  
Next, the components are stretched to their
specified axis ratios and rotated to their designated orientation
angles using the \texttt{applyShear} method.
Once all components have been built, they are added together and 
the given lensing shear and magnification are applied to the composite object, 
calling \texttt{applyShear} and \texttt{applyMagnification} respectively
(left side of \autoref{fig:flowsim}).  The \texttt{Convolve}
method is called to convolve the object with the PSF.  \galsim{}'s
\texttt{GSParams} argument can be adjusted within the \balrog{}
configuration file, to be passed as an argument to \galsim{} when
determining the target accuracy of the convolution.  \galsim{}'s
\texttt{draw} then creates an image of the simulated object.  The
\texttt{CCDNoise} method adds Poisson noise to the object's image,
setting the gain equal to that of the input image and the read noise
to zero.  Finally, the noisy object's image is assigned a center
coordinate within the original input image, and its flux is added to
the original image on a pixel-by-pixel basis.

\subsection{Measurement software}
\label{sec:measurement}

\hypertarget{hyper:config}{} The final step in the \balrog{} pipeline
is source detection and measurement.  The configuration settings of
the measurement software are an important component of this process.
Accordingly, users can pass \balrog{} any of the configuration files
\sextractor{} accepts as input and will use them to configure
\sextractor{} runs, automatically making any modifications to the
files necessary for running in the \balrog{} environment.  For
convenience, users can also override \sextractor{} settings from the
\balrog{} configuration file.

\hypertarget{hyper:nosim}{} By default, prior to inserting simulated
objects, \balrog{} runs \sex{} in \textit{association mode} over the
original image.  In this mode, we pass \sex{} a list of coordinates of
the objects to be simulated, and real objects whose positions lie
within 2~pixels\footnote{Two pixels is the \sex{} default, and
  substantially larger than our typical centroid errors.}  of any of
the \balrog{} positions are extracted into a catalog.  This allows
users to check for blending between real and \balrog{} objects, and if
preferred, remove such instances from their analyses.

\hypertarget{hyper:sim}{} Once the simulated objects are injected into
the image, \balrog{}'s default behavior makes another \sextractor{}
run in association mode, again extracting only sources whose detected
positions are within 2~pixels of one of the \balrog{} positions.  The
resulting catalog is \balrog{}'s primary output, a table of the
simulated objects' measured properties.  By running in association
mode, execution time is saved, skipping measurement of all the sources
already present in the image prior to the simulations.  This is most
relevant if the user configures \sex{} to perform measurements that
involve fitting a model to the sources, which is computationally
expensive.

We emphasize that \balrog{} is \textit{not} doing forced photometry in
association mode; we intend \balrog{} to be usable for probing
detection probability.  \sextractor{} \textit{always} runs detection
over the full image.  Measurement happens later in a separate step.
Association mode matching then decides if a detected object should be
measured or not; only detections with positions near the given
association list -- here the \balrog{} simulation positions -- are
extracted.  Association with the \balrog{} positions is why the truth
catalog enters as input to the measurement steps in
\autoref{fig:flowoverview}.

By default, \balrog{} runs in \textit{single-image mode}, meaning
simulated objects are injected into a single image, then \sex{}'s
detection and measurement are made using that same image.  \balrog{}
can also be configured to run \sextractor{} in \textit{dual-image
  mode}, where detection and measurement occur in different images.
Doing this is common in surveys; for example, DES builds a multi-band
$riz$ coadd for detection, which increases the depth of detections,
and then makes measurements in each of the passbands.  

Dual-mode \balrog{} operates slightly differently than the default single-mode.
One uses a two-call approach in order to self-consistently add the simulated objects to both images.
First one builds a detection image with simulated objects; 
this is then passed as the detection image to a subsequent \balrog{} call which adds the simulated objects
to the measurement image.  

This two-step approach to \balrog{}'s dual-mode
is a code-level choice made by the authors,
but a well-motivated one.  In the case of a multi-band detection
image, adding objects directly to the detection image is not
fundamentally correct.  One should add the \balrog{} objects to each
single-band image individually and then recoadd to build the \balrog{}
detection image; this approach most faithfully reproduces the real
data's processing.  For instance, different bands have different PSFs
and this approach convolves each separately, whereas adding to the
detection image directly would apply a single ``average'' convolution.
Accordingly, we opted to implement dual-mode as described.

\section{DES + \balrog{}}
\label{sec:sample}

Both the validation work in \autoref{sec:validation} and the
clustering measurements presented in \autoref{sec:measurements} make
use of a common sample, consisting of DES data and associated
\balrog{} simulations.  Here, we detail our data products and how they
are generated.  In \autoref{sec:inputcat}, we explain the input we
pass to \balrog{} to populate the simulation sample.  Next,
\autoref{sec:desdata} discusses the DES imaging and its processing.
\autoref{sec:desbalrog} then specifies how we configure and run
\balrog{} on this DES data.  We describe how we construct our DES and
\balrog{} catalogs in \autoref{sec:cats}, including the cuts we make
to the samples.

\subsection{Input ensemble}
\label{sec:inputcat}

Our strategy for populating simulated object parameters is to sample
magnitudes, sizes, and other \sersic{} properties from a catalog whose
probability distribution function (PDF) over the parameter space is
reasonably representative of that of the Universe on large scales.  We
begin with the COSMOS mock catalog (CMC) compiled by
\citet{jouvel2009}, who used \textsc{Le Phare} \citep{lephare2006} to fit
template spectral energy distributions to 30-band 
Cosmological Evolution Survey (COSMOS) photometry
\citep{ilbert2009}.  The template fits were convolved with the
transmission curves of several instruments, in order to generate
synthetic magnitude measurements of the COSMOS galaxies using
different cameras.  The measurements include Suprime-Cam's
\citep{suprimecam} $griz$ filter bands, comparable to DECam's $griz$
pass bands, and we adopt the Suprime-Cam magnitudes to sample our
simulation population's fluxes. At the time of the simulation, the CMC
photometry was not available for DECam's filters, but this has since
changed, and future versions of these synthetic catalogs will use the
DECam filters.

In order to assign realistic morphology to the CMC galaxies, we match
them (simple angular coordinate matching) to the morphology catalog of
\citet{great3}, consisting of single-component elliptical \sersic{}
fits to deconvolved COSMOS images.  The morphology catalog is not
complete, so we perform a nearest-neighbor four-dimensional
reweighting to the matched catalog (using 7 nearest
neighbors\footnote{This number was selected as optimal to best-match
  the CMC; we note, however, that the results of the reweighting
  method are rather insensitive to the number of nearest neighbors.}),
such that the galaxies' $griz$ magnitude distributions in the matched
catalog reproduce those of the CMC. The reweighting is analogous to
reweighting spectroscopic redshift distributions for use in
calibrating photometric redshifts, as presented in
e.g. \citet{lima2008} (and applied to DES data in
\citealt{sanchez2014}), and we will use similar methodology again in
\autoref{sec:csample}. The catalog of \sersic{} fits is for a
selection of galaxies only, and we do not reweight the CMC stars.
They are assigned to be point objects with vanishing half light radii.
In our \balrog{} simulations for this paper, we did not use the CMC
quasars, but we will include them in subsequent runs.

We make a few quality cuts prior to reweighting the galaxy sample, and
for consistency, apply the same cuts to the stellar sample where
relevant.  First, we require all three CMC colors, $g-r$, $r-i$, and
$i-z$, to be between -1 and 4.  We also reject objects whose half
light radii in the \sersic{} catalog are larger than 100\arcsec{}.
Finally, we require $i \leq 25$.  Beyond this limit, the morphology
catalog is substantially incomplete, and we lack adequate statistics
for the four-dimensional reweighting.  After applying these cuts, our
(CMC + morphology) matched catalog contains $\sim\!70,000$ objects,
and the final reweighted version of the catalog given to \balrog{}
totals $\sim\!200,000$ objects: $\sim\!190,000$~galaxies and
$\sim\!10,000$~stars.  In \autoref{sec:validation}, we find that this
catalog is of adequate size to span the parameter space used in our
analysis, and in future \balrog{} runs, we will construct the catalog
to span an even larger space.

For the purpose of this work, we populate our \balrog{} simulations by
jointly sampling brightnesses, half light radii, ellipticities,
orientation angles, and \sersic{} indexes from our reweighted
CMC~$+$~morphology-matched catalog, and simulate objects as single
component elliptical \sersic{} objects with no lensing.  The simulated
positions are randomly distributed over the celestial sphere in our
footprint, i.e. we are populating randoms which have no intrinsic
clustering.  Each object is added at the same location in the $g$,
$r$, $i$, and $z$ DES images, and drawn with the same morphology in
each band, inheriting its colors from the CMC.

\subsection{DES imaging}
\label{sec:desdata}

\begin{figure*}
	\centering
	\includegraphics[width=0.75\textwidth]{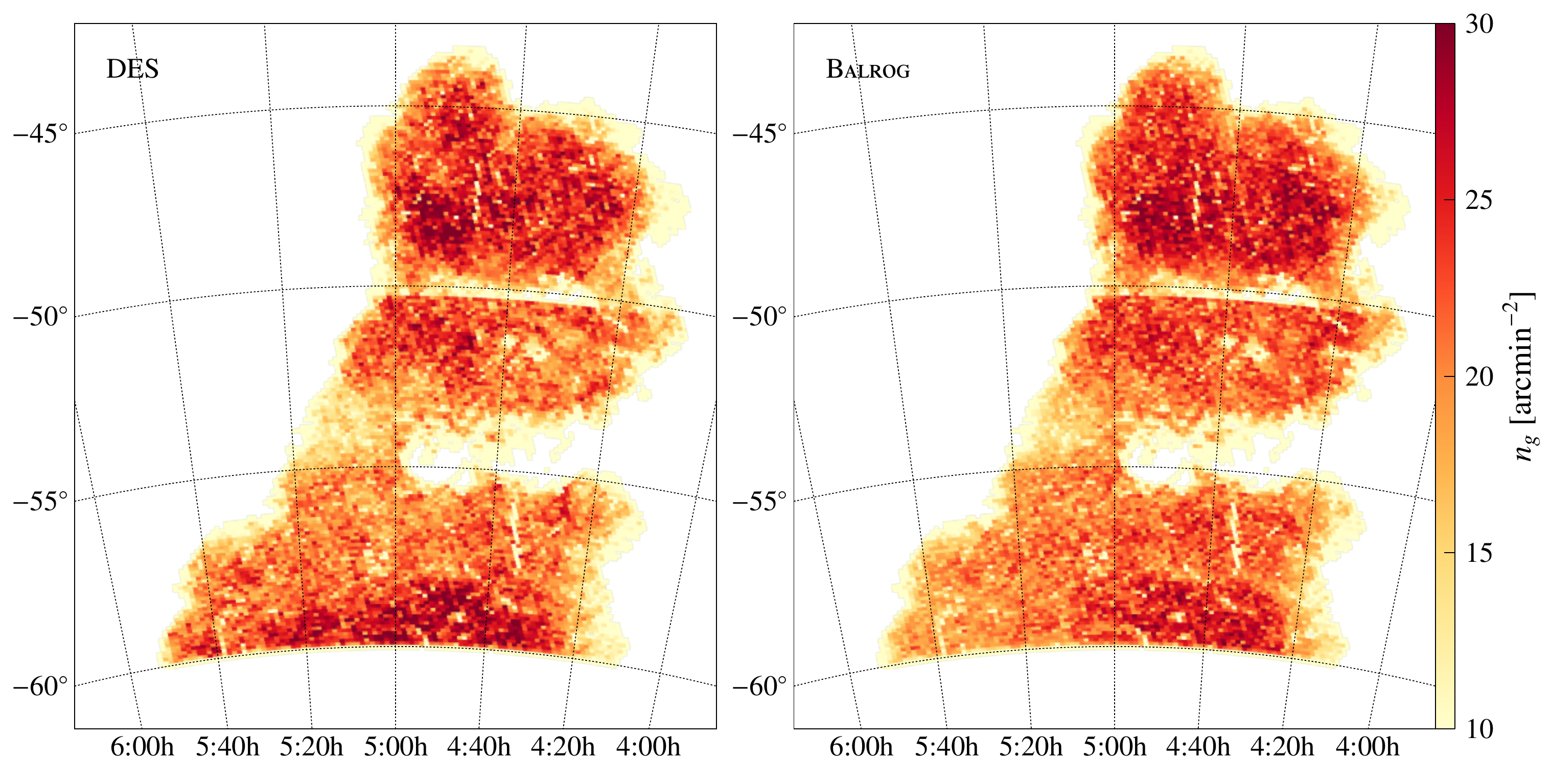}
        \caption{Map (declination vs. right ascension) of the density
          of detected DES ({\bf left}) and \balrog{} galaxies ({\bf
            right}) on the SPT-E footprint used in this
          analysis. While the two maps are very similar, there is an
          excess in counts in DES data at declination $\delta<-58$;
          this is due to increased stellar contamination caused by the
          nearby LMC.  Our \balrog{} run has made no attempt to model
          anisotropic stellar counts. }
	\label{fig:coverage}
\end{figure*}

The imaging data we consider were taken during the DES SV period,
which occurred prior to the start of first-year survey operations
\citep{diehl2014}; SV was used to verify that DECam is able to deliver
data of sufficient quality to meet DES' science goals.  We have run
\balrog{} on $178~\mathrm{deg}^2$ of the SV footprint, in an area
north of the Large Magellanic Cloud (LMC) and within the \spte{} field
-- the largest contiguous area of the SV footprint.  The \spte{} area
overlaps with the coverage of the South Pole Telescope
\citep[SPT,][]{spt-2004}, and its depth approaches that of DES
full-survey depth in some areas.  \autoref{fig:coverage} shows a map
of the detected DES and \balrog{} galaxy number density over our
selected area, where we have applied the cuts discussed in
\autoref{sec:cats}.  The following several paragraphs focus on the
processing of the DES imaging from which these samples are derived.

The DES SV data were processed through the DES Data Management (DESDM)
reduction pipeline \citep{mohr2012,desai2012}; we briefly outline
salient reductions and refer readers to the references for further
details.  First, single epoch images are overscan subtracted, a
cross-talk correction is made, and a look up table removes nonlinear
CCD responses to incident flux levels.  Bias frames are applied to
subtract out any remaining additive offsets, dome flats correct for
multiplicative variations in pixel sensitivity, and a ``star flat''
\citep[e.g.][]{manfroid1995} divides out the illumination pattern
across the detector.  Artifacts such as cosmic rays, satellite trails,
and stellar diffraction spikes are masked.  Astrometric solutions are
computed by \scamp{} \citep{scamp} matching stellar positions to the
UCAC4 reference catalog \citep{ucac4aj}.  The pipeline outputs reduced
images, along with inverse-variance weight maps and masks.

DES' photometric calibration is described in detail in
\citet{tucker2007}.  Briefly, SDSS photometric standards fields are
observed at the beginning and end of each night.  Stars from the DES
images are matched to SDSS standard stars, fitting each band's
absolute zeropoint as a linear function of airmass over all
overlapping matches.  The zeropoint for each CCD in every image is
then refit by jointly minimizing the magnitude differences between (1)
DES objects common to multiple exposures and (2) any DES objects that
match to SDSS standards.

DESDM builds coadds of the single-epoch images with \swarp{}
\citep{swarp}, using the discussed astrometric solutions and
photometric calibrations as input.  Each coadd image, known as a tile,
is $\sim\!0.5~\mathrm{deg}^2$ in area.  \swarp{} computes the
effective gain noise level of each tile as well as the combined
inverse-variance weight map. \psfex{} \citep{psfex} is then run over
the coadds to fit the PSF model, using a second-degree polynomial for
interpolation over the tile. Finally, DESDM runs \sextractor{} in
dual-image mode, using a multi-band $riz$ image for detection, to
produce the catalogs of DES objects.

The SV photometric calibration for the coadds was supplemented with
stellar-locus regression (SLR), which uses the near universality of
the colors of Milky Way halo stars as a means to fit for photometric
zeropoints \citep[e.g.][]{high2009}.  Our SLR corrections \eliipp{}
were implemented with a modified version of the \textsc{big-macs}
stellar-locus fitting code \citep{kelly2014}.  All corrections were
made relative to an empirical reference locus derived from calibrated
standard stars observed on a photometric night.  We recompute coadd
zeropoints over the full SV footprint on a HEALPix \citep{healpix}
grid of \texttt{NSIDE = 256}, using bilinear interpolation to correct
all objects in the catalog at a scale of better than
$\sim\!14\arcmin$.  We use $J$ band magnitudes from the Two Micron All
Sky Survey (2MASS) stellar catalog \citep{twomass} as an absolute
calibration reference, which yields absolute calibration uniformity of
better than $2\%$, with color uniformity $\sim\!1\%$.

\subsection{Running \balrog{}}
\label{sec:desbalrog}

The input we give to \balrog{} is made up of the data products
discussed in the previous section: the coadded SV images from DESDM,
as well as their inverse-variance weight maps, PSF models, astrometry,
photometric zeropoints, and effective gains.  We self-consistently add
the same \balrog{} objects to the $g$, $r$, $i$, and $z$ images, build
an $riz$ detection image for each realization using identical \swarp{}
configuration as DESDM, and then run \balrog{} over each band with
\sex{} configurations, which again match those of DESDM.

We make use of the SLR offsets introduced in \autoref{sec:desdata} in
our imaging simulations.  We employ \balrog{}'s user-defined function
API to read the SLR zeropoints and make position-dependent
modifications to the simulated fluxes in each image, in addition the
usual single zeropoint used by \balrog{}.  This takes an input truth
magnitude and adjusts it back to the pre-SLR flux scale, i.e. the
original calibration for the coadd images.

In each \balrog{} realization we add only $1,000$~objects to the image
(of area $\sim\!0.5~\mathrm{deg}^2$), in order to keep the
\balrog{}-\balrog{} blending rate low.  We iterate each coadd tile
100~times, simulating a total of 100,000~objects per DES coadd
tile. Combining the results generates a \balrog{} output measurement
catalog which is approximately the same size as the DES measurement
catalog.  The total run time for our \balrog{} simulations was
approximately $30,000$~CPU-hrs, much less than the time needed by
DESDM to process the data.

Admittedly, injecting our \balrog{} objects directly into the coadds
instead of self-consistently into each overlapping single-epoch image
is less ideal. For example, the coadd PSF is not as reliable of a
model of the data as is simultaneously using the full set of
single-epoch PSFs. However, the single-epoch version of \balrog{} is
roughly ten times more computationally expensive, and we opt to test
the simpler approach first.  Using \balrog{} in other DES analyses
which are more sensitive to the PSF and which directly use
single-epoch level information (such as weak lensing ones) will
require running on all the single-epoch images. In this work, our
measurements are focused on galaxy clustering, and we demonstrate that
the coadd approximation is sufficient in this context.

\subsection{Catalog selection}
\label{sec:cats}

%
%

\begin{table*}
\caption{\texttt{MODEST\_CLASS} selection.}
\label{tab:modest}
\begin{tabular}{l l}
\hline
Galaxies & Stars \\
\hline
\texttt{(FLAGS\_I $\le 3$) AND NOT} &  \texttt{(FLAGS\_I $\le 3$) AND} \\
\texttt{( \hspace*{0.03cm} ((CLASS\_STAR\_I $> 0.3$) AND (MAG\_AUTO\_I $< 18$))} &  \texttt{( \hspace*{0.15cm} (CLASS\_STAR\_I $> 0.3$)} \\
\texttt{OR ((SPREAD\_MODEL\_I + 3*SPREADERR\_MODEL\_I) $< 0.003$)} & \texttt{AND (MAG\_AUTO\_I $< 18$)} \\
\texttt{OR ((MAG\_PSF\_I $> 30.0$) AND (MAG\_AUTO\_I $< 21.0$))} & \texttt{AND (MAG\_PSF\_I $< 30.0$)}\\
\texttt{)} & \texttt{OR \hspace*{0.0cm} (((SPREAD\_MODEL\_I + 3*SPREADERR\_MODEL\_I) $< 0.003$) AND} \\
& \texttt{\hspace*{0.42cm} ((SPREAD\_MODEL\_I + 3*SPREADERR\_MODEL\_I) $> -0.003$)))} \\
& \texttt{)} \\
\hline
\end{tabular}
\end{table*}

%


To construct the DES sample, we download the SV coadd data from the
DESDM database of \sex{} measurements, returning detections from the
same areas where \balrog{} was run. We then apply the SLR zeropoint
shifts to both the DES and the \balrog{} catalogs. At this point, the
full \balrog{} and DES catalogs total $\sim\!16$~million detections
each.

Next, we apply some quality cuts to both samples.  In
\autoref{sec:measurements}, we undertake galaxy clustering
measurements, and the quality cuts we make are similar to ones made
in the \textit{benchmark} DES clustering analysis
of \citet{benchmark}.
We base our cuts on a subset of their selection
criteria as means to help achieve a reasonably well-behaved source
population.

First is a simple color selection:%
\footnote{\citet{benchmark} use \texttt{DETMODEL} colors, but we choose to use \texttt{AUTO} colors.}
\begin{align*}
-1 &< \texttt{MAG\_AUTO\_G} - \texttt{MAG\_AUTO\_R} < 3 \; \\
 \texttt{AND} \; \; -1 &< \texttt{MAG\_AUTO\_R} - \texttt{MAG\_AUTO\_I} < 2 \; \\
 \texttt{AND} \; \; -1 &< \texttt{MAG\_AUTO\_I} - \texttt{MAG\_AUTO\_Z} < 2.
\end{align*}
This helps to eliminate objects inside regions which are contaminated in one filter band's
image, but not the others, such as satellite or airplane trails.

Furthermore, we make a cut based on \sex{} position measurements.
Among the \sex{} detections, there exists a class of objects whose
windowed centroid measurements are significantly offset in different
filter bands,\footnote{We suggest \textit{astrometric color} as the
  name for this effect.}  up to over a degree in the worst cases.
This is to be expected for objects with low signal-to-noise ratios,
since detection occurs in $riz$, while measurement occurs in each band
independently, and the centroid measurement for a dropout in a given
band is essentially unconstrained. However, large positional offsets
persist at all signal-to-noise levels, such that about 2\% of all
objects at any signal-to-noise have significant offsets.  We reject
any object with large ($> 1$\arcsec{}) offset between the $g$- and
$i$-band centroids, which has been detected with $> 5\sigma$
significance in $g$-band.

We also apply the mask used by \citet{benchmark}.  (Specifically, we
use the mask as it exists prior to introducing redshift dependence.)
The details of
the mask's construction are found in \autoref{sec:mask}; in brief, it
is based on five criteria:
\begin{enumerate}
\item coordinate cuts to select \spte{} area north of the LMC,
\item excising regions with the highest density of large positional offset objects discussed above,
\item removing objects in close proximity to bright stars,
\item selecting regions with $10\sigma$-limiting magnitude of $i > 22.5$, and
\item requiring detections over a significant fraction of the local area.
\end{enumerate}

The cuts we have mentioned in this section are not strictly necessary
for the validation tests presented in \autoref{sec:validation} to
follow. In fact, \balrog{} is able to populate objects like the ones
that have been cut into the simulated sample. However, we are most
interested in \balrog{}'s behavior for objects which will survive into
a science analysis. Therefore, we choose to exclude them from the
clustering study presented in \autoref{sec:measurements}.

Throughout the remainder of our analysis, we also remove any objects
from the \balrog{} simulation catalog which have a matched counterpart
in the catalog generated by running \sextractor{} prior to inserting
any simulated objects (cf. \autoref{sec:measurement}). Doing so
removes approximately 1\% of the \balrog{} catalog.  Some of these
objects are genuine \balrog{} objects, some are DES objects, and
others are blends of the two, depending on the relative brightness of
the input \balrog{} object compared to the DES object found in the
image at the simulation location.  This choice does have a small 
impact ($\sim \! 1\%$) on 
the clustering: including the ambiguous matches effectively mixes some
real galaxies into the randoms used for clustering, artificially
suppressing the clustering signal; excluding the ambiguous matches has
the opposite effect. We discuss this issue along with other
fundamental limitations of the embedding simulation approach in
\autoref{sec:caveats}.

The final selection mechanism we use is \sg{} separation.  \Sg{}
separation is accomplished with the \texttt{MODEST\_CLASS} classifier,
which is explained in e.g. \citet{ufig-des}, and utilized in
additional DES analyses such as \citet{sva1-mass-maps-mnras} and
\citet{boris_maps}.\footnote{As noted in \autoref{sec:bscomp},
  \citet{benchmark} use a new quantity -- \texttt{WAVG\_SPREAD\_MODEL}
  -- for \sg{} separation.}  The classifier has been tested with DES
imaging of COSMOS fields. \autoref{tab:modest} lists the full
\texttt{MODEST\_CLASS} selection criteria. It incorporates \sex{}'s
default \sg{} classifier \texttt{CLASS\_STAR}, which is based on a
pre-trained neural network, as well as morphological information about
how well the object resembles the PSF; for each object,
\texttt{SPREAD\_MODEL} measures a normalized linear discriminant 
between the best-fit local PSF model derived with \psfex{},
and a slightly more extended model made from the PSF
convolved with a circular exponential disk
\citep[see e.g.][]{desai2012, bouy2013, soumagnac2015}.  \texttt{SPREADERR\_MODEL} is the error
estimate for the \texttt{SPREAD\_MODEL} measurement.

Including the cut on \texttt{SPREADERR\_MODEL}, in addition to
\texttt{SPREAD\_MODEL} alone, improves the faint end galaxy
completeness. Including the \texttt{MAG\_PSF} cut improves the purity
at the bright end. \citet{soumagnac2015} investigate more
sophisticated means of \sg{} separation, such as machine learning
techniques beyond \sex{}'s pre-trained \texttt{CLASS\_STAR}, and in a
subsequent publication \jelenaip{}, we will present a neural network
approach trained on \balrog{} data. In \autoref{sec:contam} we
demonstrate that \texttt{MODEST\_CLASS} suffices for our current
analysis.

After applying all the cuts discussed in this section, the DES and
\balrog{} galaxy catalogs total $\sim\!10$~million objects each.
These are the samples whose number densities we mapped in
\autoref{fig:coverage}. We use these catalogs as our primary data
products in \autoref{sec:validation} and \autoref{sec:measurements}.

\section{\balrog{} Validation}
\label{sec:validation}

\begin{figure*}
	\centering
	\includegraphics[width=0.99\textwidth]{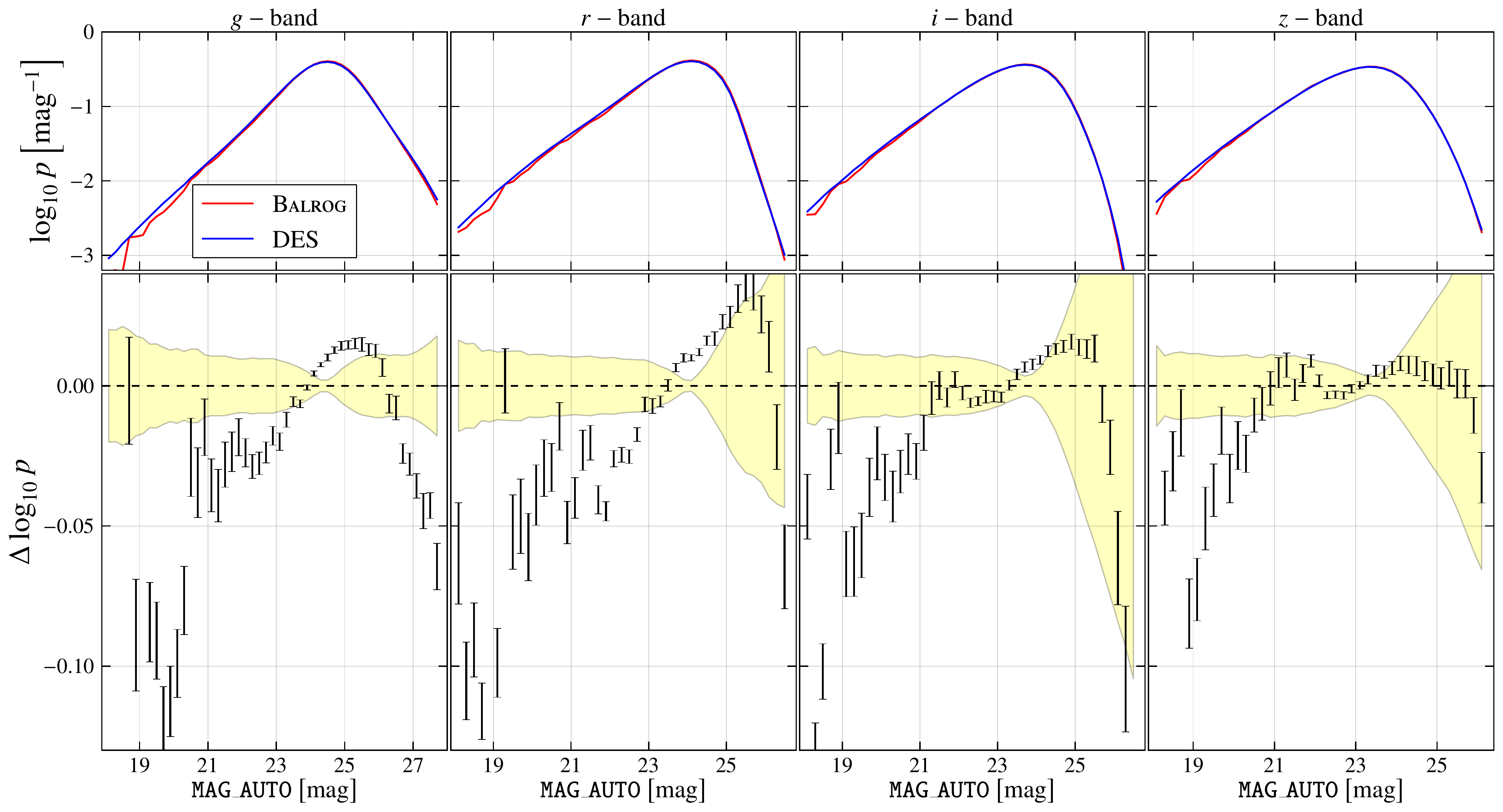}
	\caption{{\bf Top: }Magnitude distributions (PDFs) of DES and \balrog{} galaxies
          in four DES filters. {\bf Bottom:} Illustration of the
          difference between DES and \balrog{} magnitude
          distributions is shown in black; errors are estimated
          from jackknife resampling, as described in \autoref{sec:jk}. The yellow band shows the
          sample variance of the DES catalogs, also \jk{} estimated.}
	\label{fig:mags}
\end{figure*}

\begin{figure*}
	\centering
	\includegraphics[width=0.99\textwidth]{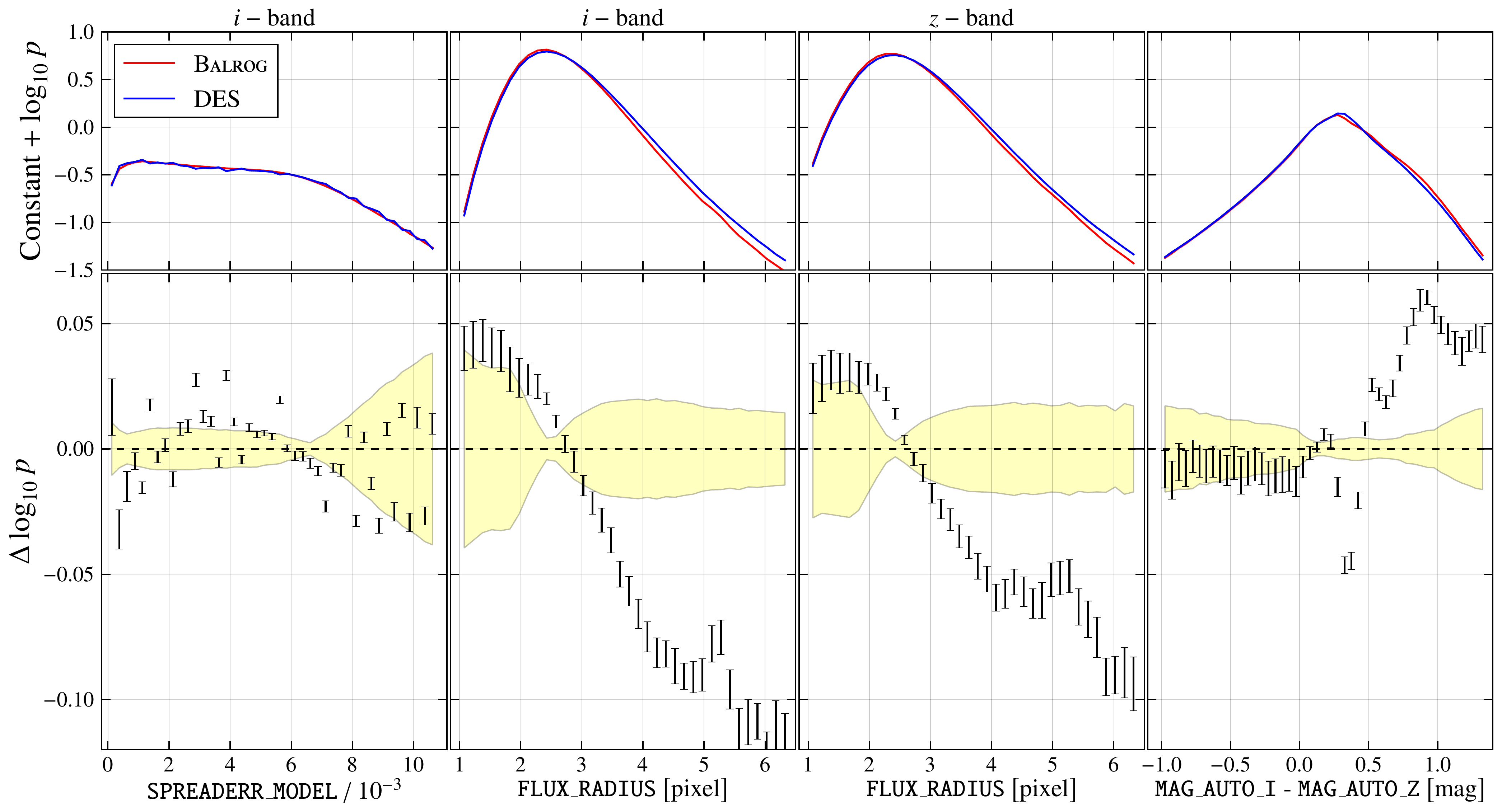}
	\caption{{\bf Top:} An idiosyncratic selection of measured
          photometric properties. The logarithmic PDFs for DES and \balrog{} in each panel have been
	  shifted by an additive constant.
	  {\bf From left to right:} reported errors in one of
          \sex{}'s stellarity measures, 
          $i-$band size, $z-$band size, and $i-z$ color.
	  We expect the filter
          mismatch described in \autoref{sec:inputcat} to drive at least
          some of the color residuals. 
	  Cosmic variance in the COSMOS
          field is also present, though we have made no rigorous
          attempt to estimate its impact here.
	  {\bf Bottom:} analogous to \autoref{fig:mags};
	  in black, we show the difference between the DES and \balrog{}
          distributions in the top panel.  The yellow band indicates
          the sample variance of the DES measurements.
	  All errors are estimated from \jk{} resampling.
	  (See \autoref{sec:jk} for further details.) }
	\label{fig:other}
\end{figure*}

To validate \balrog{}'s functionality, we analyze the catalogs
constructed in \autoref{sec:cats}, testing if the properties of the
\balrog{} objects are representative of the DES data. For our
\balrog{} runs, we have attempted to build an input catalog which is
deeper than our actual DES data.  If this input distribution is indeed
an adequately representative sample, and our DES calibrations (PSF,
flux calibration, etc.) are well measured, running the simulations
through \balrog{} should successfully reproduce measurable properties
of the DES catalogs.

The \balrog{} and DES comparison tests presented in this section are
as follows: \autoref{sec:1d} plots one-dimensional distributions of
measured \sex{} quantities, \autoref{sec:2d} does similarly for
two-dimensional distributions, and \autoref{sec:nboris} considers
number density fluctuations.  \autoref{sec:2d} and
\autoref{sec:nboris} include assessments of the populations' behavior
as a function of observing conditions of the survey.  The one- and
two-dimensional distributions offer a general overview of the
agreement between \balrog{} and DES, and the number density tests
validate that the agreement is sufficient to use our \balrog{}
galaxies as randoms in \autoref{sec:measurements}'s clustering
measurements.

We also make note of \autoref{sec:jk}, where we explain our
jackknifing procedure, used to estimate errors in this section, as
well as in \autoref{sec:measurements}.  To summarize, we use a
$k$-means algorithm to separate our data sample into 24~spatial
regions of roughly equal cardinality, then leave one region out in
each \jk{} realization and calculate the covariance over the
realizations.

\subsection{One-dimensional distributions}
\label{sec:1d}

We compare the $griz$ magnitude (\texttt{MAG\_AUTO}) distributions of
galaxies, for both the DES and the \balrog{} samples in
\autoref{fig:mags}. The top row of the figure plots each band's
$\log_{10} p$, the logarithm of the PDF, and the second row plots the
difference in this quantity between \balrog{} and DES, i.e. the
fractional deviation between the two PDFs. The error bars plotted are
the square root of the diagonal elements of the \jk{} covariance
matrix, as described in \autoref{sec:jk}, where we have jackknifed the
difference curve, $\Delta \log_{10} p$.  For \texttt{MAG\_AUTO} $
\gtrsim 21$ -- the region of the parameter space occupying the bulk of
the galaxies -- \balrog{} reproduces the DES distribution to better
than 5\% differences, approaching 1\% over some intervals.  The yellow
bands in bottom row of \autoref{fig:mags} show the \jk{} errors of the
DES PDFs plotted in the top row.  In the densest parameter space
regions, many of data points of the differences between \balrog{} and
DES are within the DES variance, particularly in the $i$ and
$z$-bands.  This means that in these regions of magnitude space,
\balrog{} galaxies are statistically indistinguishable from DES
galaxies.

We also make plots analogous to \autoref{fig:mags}, using measured
quantities other than single-band magnitudes (\autoref{fig:other}).
In each of the top panels, we have shifted $\log_{10} p$ for both the
DES and \balrog{} curves by an additive constant, so all the panels
share a similar range on the $y$-axis.  We plot distributions in
(\texttt{MAG\_AUTO\_I} - \texttt{MAG\_AUTO\_Z}) color, $i$-band
\texttt{SPREADERR\_MODEL}, as well as $i$- and $z$-band
\texttt{FLUX\_RADIUS}.  \texttt{FLUX\_RADIUS} measures the PSF
convolved half light radius.  \texttt{SPREADERR\_MODEL} is the error
in the \texttt{SPREAD\_MODEL} measurement introduced in
\autoref{sec:desdata}.  We again find that \balrog{} reproduces DES to
$\sim$$\,5\%$ differences or better in the bulk of the distributions;
this result holds across bands and across different \sex{} quantities.
We chose to include \texttt{SPREADERR\_MODEL} in our comparison
because it is not obviously straightforward to simulate directly; it
is the error in a measurement unique to \sex{}.  Nevertheless,
\balrog{} is able to recover a distribution similar to DES.

\begin{figure}
	\centering
	\includegraphics[width=0.4\textwidth]{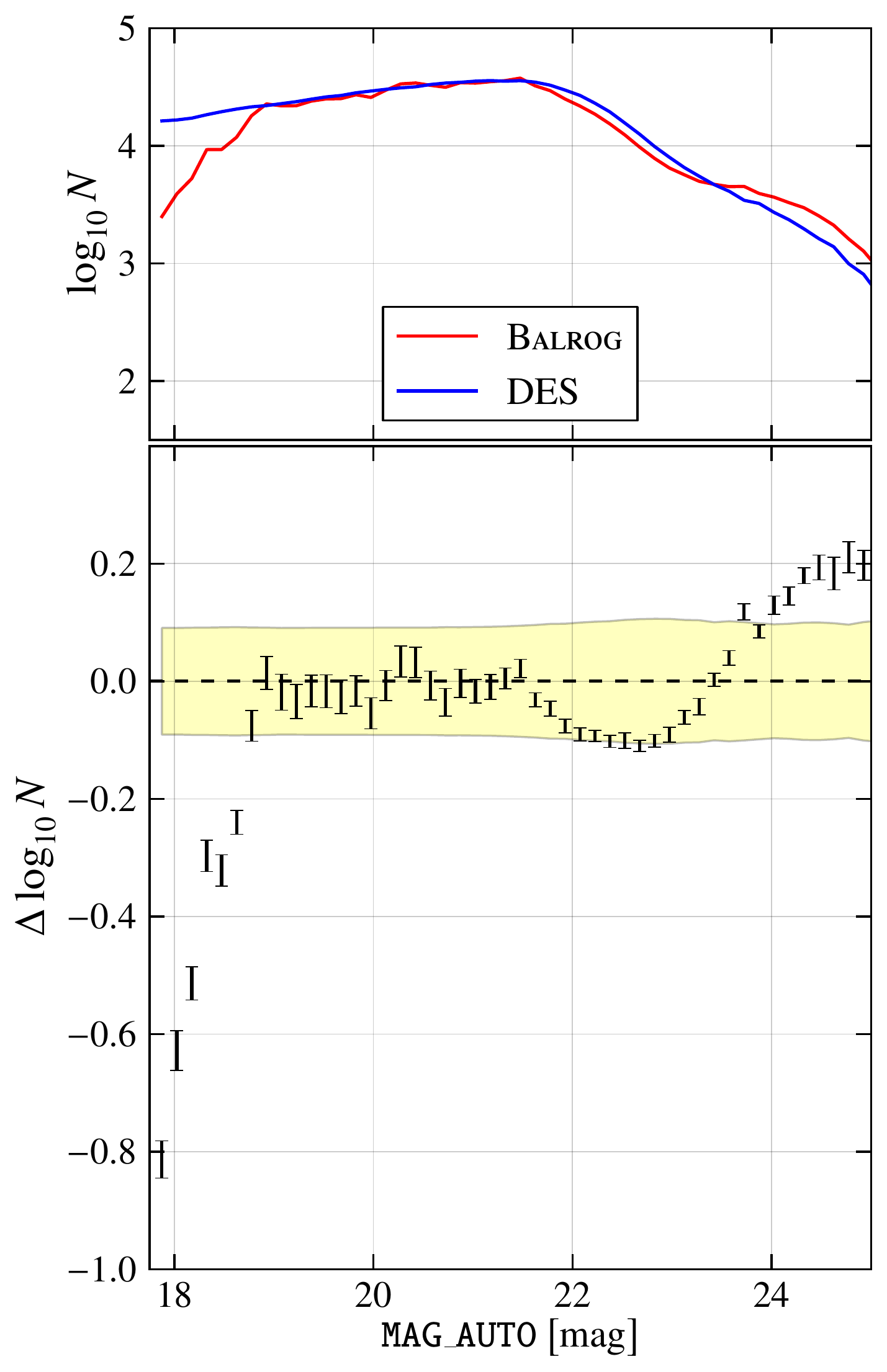}
	\caption{Stellar magnitude distributions in DES and \balrog{},
          $i$-band.  The \balrog{} curve has been normalized by
          selecting $23 < \texttt{MAG\_AUTO\_I} < 24$ objects, and
          multiplying by the detected DES star-to-galaxy number ratio
          divided by the detected \balrog{} star-to-galaxy number
          ratio (as in \autoref{sec:contam} when estimating DES
          stellar contamination levels).  At the bright end, the
          difference is primarily a result of the lack of bright stars
          ($i<19$) in the CMC catalog 
	  (due to saturation in the COSMOS images) 
          used to seed the \balrog{}
          simulations.  
          Furthermore, 
          the stellar density varies substantially
          across the SV field (see \autoref{fig:regions}), so the
          COSMOS stellar population is not necessarily
          representative.}
	\label{fig:star-mags}
\end{figure}

\begin{figure*}
	\centering
	\includegraphics[width=0.75\textwidth]{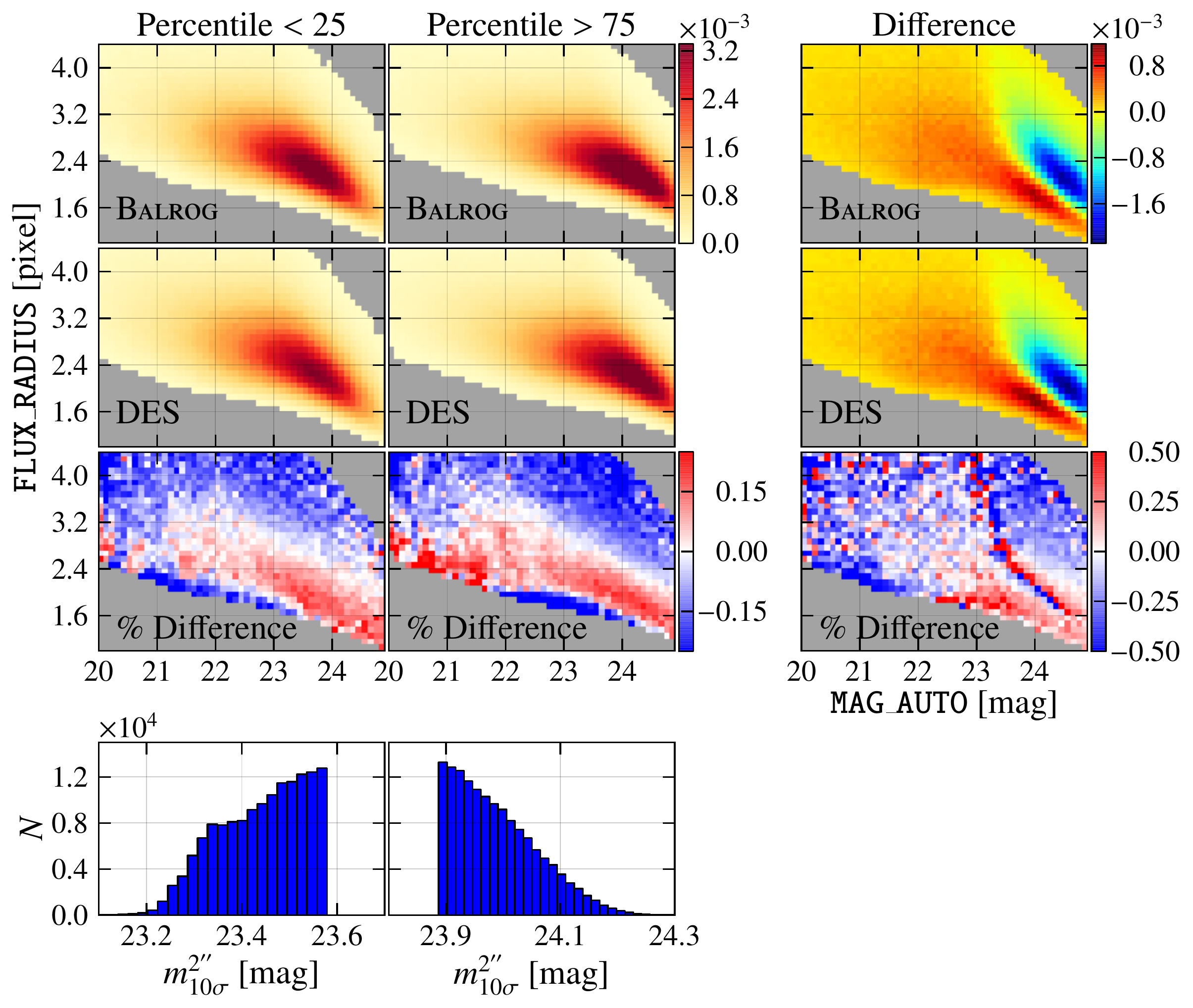}
	\caption{{\bf Top left block:} These six panels show the
          sensitivity of the $i$-band size-magnitude distributions of
          DES and \balrog{} galaxies to survey depth. The color scale
          in the upper four panels shows normalized counts.  {\bf
            Bottom block:} Histograms of the depth selection in
          each column; we split into two samples: the deepest 25\% of
          the area and the shallowest 25\% of the area.  {\bf Right
            block:} Differences between the left panels. The bottom
          panel shows the difference between the above two
          differences. While these histograms are noisy, this figure
          shows that \balrog{} well captures the effect of depth on
          the measured galaxy properties. The systematic differences
          visible here are mainly due to the small differences between
          the DES and CMC catalogs.}
	\label{fig:2d-maglim}
\end{figure*}

If \balrog{} were a perfect model of the data, $\Delta \log_{10} p$
would be consistent with zero everywhere, but in practice, we do not
expect to recover this result.  Even in the limit of perfect survey
calibrations (PSF, photometric calibration, etc.), one would need a
completely representative input population to recover perfect
agreement.  We have made the assumption that single component
elliptical \sersic{} objects fully describe the galaxy population, but
this is not strictly true.  Moreover, COSMOS (point) sources begin
saturating for $i \lesssim 19$ \citep{acs-data, capak2007}.  The CMC
does not include such objects, and thus our reweighted catalog is not
expected to be entirely complete at bright magnitudes.  Furthermore,
COSMOS is a small field ($\sim\!2~\mathrm{deg}^2$): with limited
statistics and cosmic variance, it is not necessarily entirely
representative of a larger area survey like DES, especially at
brighter and larger size limits; this could be another contributing
factor why \balrog{}'s brighter and larger galaxies are less
representative of DES than its fainter and smaller ones.  Finally, we
have also used Subaru filters for our input magnitudes, (because DECam
ones were not available), which will introduce some error when
comparing \balrog{} and DES distributions.

\autoref{fig:mags} and \autoref{fig:other} plotted galaxy selections,
but our \balrog{} run also included stars.  \autoref{fig:star-mags}
shows the $i$-band DES and \balrog{} stellar distributions.
We have normalized the \balrog{} curve in the top panel in the following way:
$N$ in each bin of the \balrog{} curve is multiplied by the detected
star-to-galaxy number ratio in DES divided by the detected
star-to-galaxy number ratio in \balrog{}, where we have selected
detections from $23 < \texttt{MAG\_AUTO\_I} < 24$.  (This is the same
way we normalize when estimating the DES stellar contamination ratio
of our faint clustering sample in \autoref{sec:contam}.) 

There is more variation in the stellar distributions compared to the
galaxy distributions, and this is to be expected.  First, we see a
large deficit due to the effects of saturation in the COSMOS imaging
at $i \lesssim 19$, as mentioned above. Stars are more compact than
galaxies and thus more heavily affected by saturation.  Furthermore,
the stellar population intrinsically fluctuates much more strongly
across the sky than the galaxy population, and the small stellar
sample from the COSMOS field need not be entirely representative of
DES as a whole.  Indeed, the DES catalog contains more detected stars
than the \balrog{} catalog.%
\footnote{$\sim \! 35\%$ more, with increased deviation near the
  LMC.}%
For this analysis, we are primarily interested in galaxies and the
COSMOS stellar population suffices; however, in a broader context, we
offer it as an example of how one should be mindful to use \balrog{}
with an input simulation population which is appropriate for one's
science case.

\subsection{Two-dimensional distributions and observing conditions}
\label{sec:2d}

\begin{figure*}
	\centering
	\includegraphics[width=0.75\textwidth]{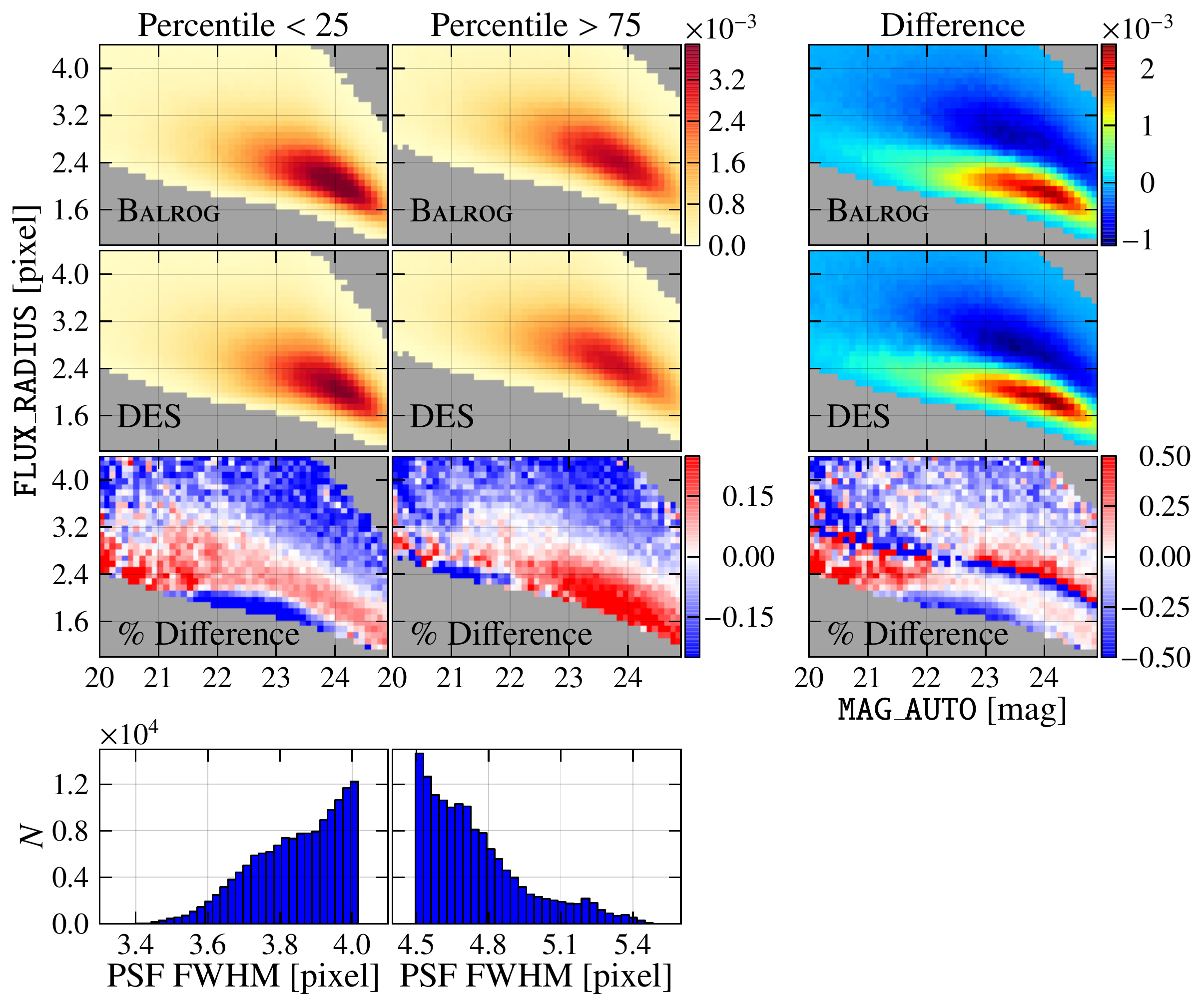}
	\caption{Analogous to \autoref{fig:2d-maglim}, but instead
          showing the ($i$-band size-magnitude) dependence on average
          seeing in the coadd images. Again, \balrog{} successfully
          captures the dependence of the measured galaxy properties on
          observing conditions. The systematic differences visible
          here are mainly due to the small differences between the DES
          and CMC catalogs.}
	\label{fig:2d-fwhm}
\end{figure*}

In addition to validating \balrog{}'s ability to recover DES'
distributions of measured quantities, we also need to test if
\balrog{} behaves like DES as a function of observing properties of
the survey. \citet{boris_maps} have constructed HEALPix maps of
several characteristics of the DES SV observations, including PSF full
width at half maximum (FWHM), 10$\sigma$ limiting magnitude in
2\arcsec{} apertures ($m_{10\sigma}^{2^{\prime
    \prime}}$),\footnote{These measurements are analogous to the
  \texttt{MANGLE} depths (discussed in \autoref{sec:mask}), without
  quite as fine a resolution.} airmass, sky brightness, and sky
variance (where the square root of sky variance is called sky
$\sigma$).  Each map computes an average of a given quantity in the
overlapping single-epoch observations for any pixel in the map, using
either an ordinary mean or a weighted mean, where the weights are
taken from the single-epoch inverse variance maps. We use the maps of
\citet{boris_maps}, available at a resolution of $\texttt{NSIDE} =
4096$, and compare \balrog{}'s behavior against DES' behavior as a
function of the observing conditions.

First, we split our DES and \balrog{} galaxy samples into two
divisions according to the local 10$\sigma$ magnitude limit, selecting
the top and bottom 25 percentiles.  The depth histograms for these two
samples are shown in the bottom row of \autoref{fig:2d-maglim}, with
the shallower sample in the left column. The first two rows of this
leftmost column show normalized \balrog{} and DES two-dimensional
histograms in the $i$-band size-magnitude plane for the shallower
magnitude limit selection. The top two rows of the middle column show
likewise for the deeper selection. The third row quantifies the
fractional difference between the \balrog{} and DES rows. Like the
one-dimensional examples, in the densest regions of parameter space
\balrog{} and DES largely agree.  Moreover, simultaneous agreement in
both depth samples offers evidence that \balrog{} traces the
distribution's properties as a function of magnitude limit.  The
rightmost column of \autoref{fig:2d-maglim} further tests this: the
top two rows in this column plot the \balrog{} and DES differences of
the shallower and deeper distributions, and the third row plots the
fractional difference between the two rows above, i.e. this panel
compares the DES and \balrog{} magnitude-size derivative with respect
to magnitude limit.  Except in regions of sharp change, agreement in
well-sampled areas of parameter space is typically $\sim\!10\%$
differences, offering additional evidence that \balrog{} reasonably
tracks the DES changes with observing conditions.

We have made analogous plots to \autoref{fig:2d-maglim}, splitting on
properties other than magnitude limit, and find similar results.
\autoref{fig:2d-fwhm} offers another example, dividing the sample
based on PSF FWHM.  The figure is largely reminiscent of
\autoref{fig:2d-maglim}.

\begin{figure*}
	\centering
	\includegraphics[width=0.99\textwidth]{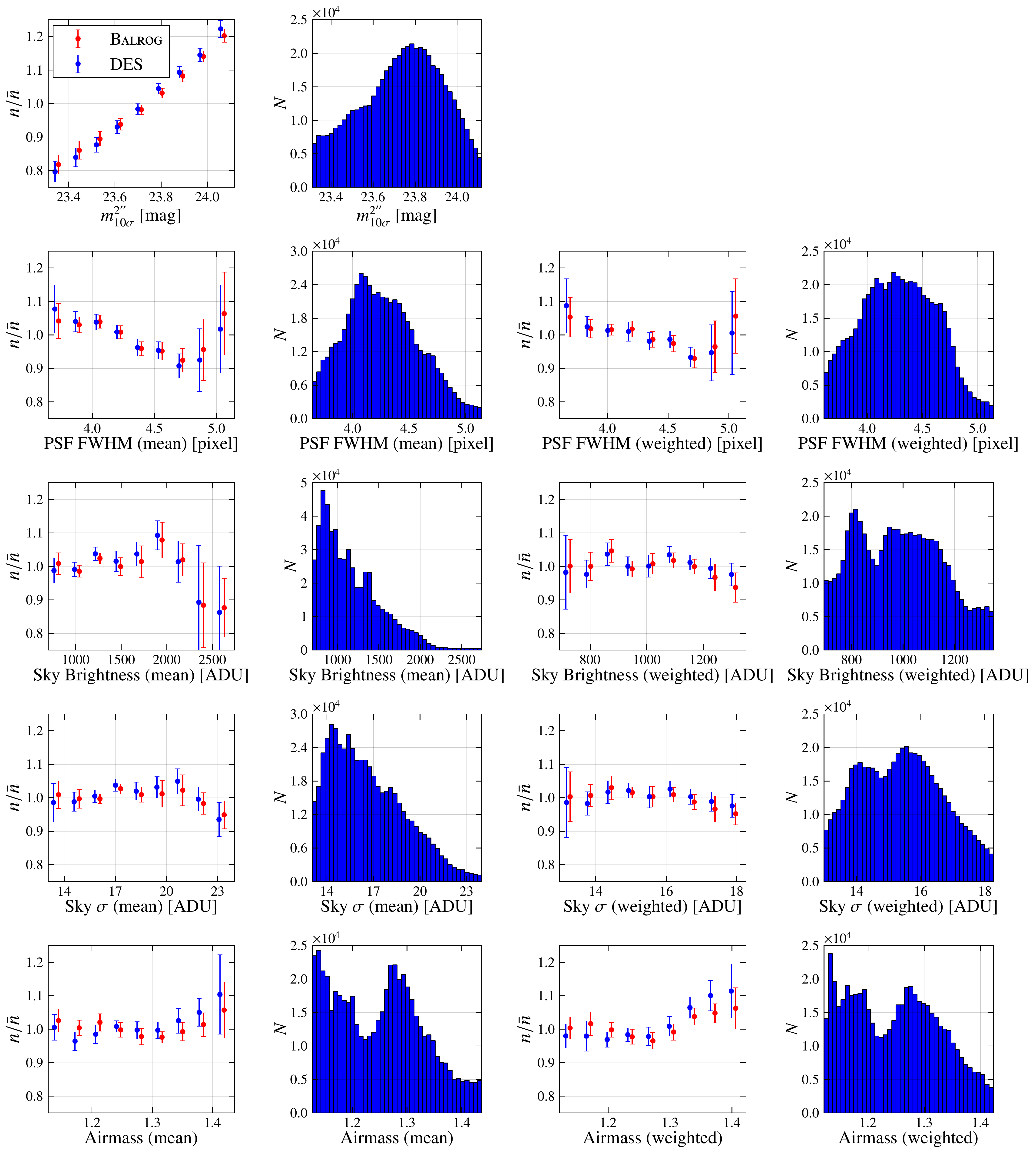}
	\caption{Number density fluctuations in the
DES and \balrog{} galaxy samples as a function of $i$-band survey
properties, binning the survey properties over the 2 to 98~percentile range. 
(The DES and \balrog{} curves have been slightly offset for visual clarity.)
For each number density bin, we count the number of
galaxies in the given pixels, divide by the area covered by those
pixels, and normalize by the average density over the full sample. 
Error bars are estimated from 24 \jk{} resamplings of the curves (cf. \autoref{sec:jk}).
Alongside the number density plots are 
histograms of the survey observing conditions, again binned over the
2 to 98 percentile range.}
	\label{fig:boris-full}
\end{figure*}

\subsection{Number density and observing conditions}
\label{sec:nboris}
To conclude this section, we test \balrog{}'s ability to recover
DES-like number density fluctuations as a function of the survey
properties mapped by \citet{boris_maps}, i.e. we investigate if
\balrog{} recovers DES' window function over the observing
conditions. If this check is successful, it means \balrog{} galaxies
can be used as a set of random points in a clustering analysis in
order to correct for varying detection probability over the
footprint. In \autoref{sec:1d} and \autoref{sec:2d}, we demonstrated
that \balrog{} is largely, but not perfectly, representative of the
DES data; assessing whether or not agreement is \textit{good enough}
depends on one's science case.  Here, we investigate if the agreement
is at an adequate level such that \balrog{} detection rates are
representative of DES detection rates, within the respective error
estimates.

\autoref{fig:boris-full} plots number density fluctuations in our full
DES and \balrog{} galaxy samples as a function of $i$-band survey
properties, binning in each survey property over the 2 to
98~percentile range. Alongside these number density plots, we also
include the histograms of the survey observing conditions over the
same range. For each number density bin, we count the number of
galaxies in the given pixels, divide by the area covered by those
pixels, and normalize by the average density over the full sample. We
plot both the DES and \balrog{} samples, where the points have been
slightly offset for visual clarity. The error bars on each set of
points are estimated by 24 \jk{} realizations of the curve, as
described in \autoref{sec:jk}. We find that the DES and \balrog{}
results are consistent with each other within the errors estimates,
which demonstrates \balrog{}'s modeling as adequate to recover the DES
window function over the tested sample. We have repeated this exercise
using the survey properties across other filter bands, finding
consistent results.

\section{Angular Clustering}
\label{sec:measurements}

The final test of the \balrog{} catalogs described in this paper is
their use in systematic error amelioration for an angular clustering
measurement. Selecting the \balrog{} catalog in the same way as the
real catalog produces a sample with a nearly identical window function
as the data's.  The \balrog{} catalogs have inherited systematic
errors in the imaging and analysis pipelines, but otherwise have no
intrinsic clustering themselves.  Hence, using them as randoms in a
two-point estimator is a simple and efficient way of removing the
systematic errors while maintaining the real clustering signal. The
rest of this section describes how this is done.

We describe what we believe are the practical and fundamental
limitations of embedding simulations for clustering measurements in
\autoref{sec:caveats}. \autoref{sec:est-alg} discusses the algorithms
we use to make our $w(\theta)$ measurements.  In
\autoref{sec:DESsample}, we select two magnitude-limited DES samples
and perform tests in \autoref{sec:contam} to show that stellar
contamination is unimportant for the measured angular clustering
signals. In \autoref{sec:csample}, we select similar populations from
the public COSMOS galaxy catalog of \citet{capak2007}
($\sim\!2~\mathrm{deg}^2$ in area) and match them to our DES
samples. \autoref{sec:wmeas} then demonstrates that over the
measurable range of angular separations, our \balrog{}-corrected DES
measurements reproduce the much deeper COSMOS measurements, but with
substantially improved accuracy and range, owing to the larger survey
volume. The shapes of our $w(\theta)$ results follow model predictions.

\subsection{Caveat Likelihood}
\label{sec:caveats}
Fundamentally, our simulated galaxies are sampling the likelihood
function that connects the measured parameters
($\boldsymbol{\alpha}_{\rm meas}$) of stars and galaxies to the
underlying true parameters ($\boldsymbol{\alpha}_t$) of objects in the
DES images. In general, the detection probability and measurement
biases for some particular galaxy depend on the rest of the galaxies
in the image, even including objects that may not be
detected. Denoting the set of all relevant object parameters by
$\{\boldsymbol{\alpha}\}$, and expressing the dependence on position
on the sky $\boldsymbol{\theta}$ explicitly, we can write:
\begin{align}
\mathcal{L} = p(\boldsymbol{\alpha}_{\rm meas}\: | \: \left\{\boldsymbol{\alpha}_t\right\}, \boldsymbol{\theta}).
\end{align}
$\mathcal{L}$ is meant to incorporate sample selection criteria, so
the probability $p(\boldsymbol{\theta})$ of {\it any} object being
selected for analysis is the likelihood integrated over the true and
observed properties:
\begin{align}
p(\boldsymbol{\theta})= \int\:  p(\boldsymbol{\alpha}_{\rm meas}\: | \:
\left\{\boldsymbol{\alpha}_t\right\}, \boldsymbol{\theta})\:d\boldsymbol{\alpha}_{\rm
  meas} d\{\boldsymbol{\alpha}_t\}.
\label{eqn:window}
\end{align}
This is also sometimes called the window function, and it is this
function that the random catalogs used in correlation function
estimators (like \autoref{eq:ls}) are meant to be sampling.

The likelihood sampled by the \balrog{} catalogs is only an
approximation of the true $\mathcal{L}$. In part, this is a result of
simplifications made in the simulation. Our input catalog, for
instance, is limited in its realism by the galaxy templates used to
generate the synthetic colors in the COSMOS mock catalogs and by the
finite size of the COSMOS field. This limitation is equivalent to
integrating in \autoref{eqn:window} only over the regions of
$\boldsymbol{\alpha_t}$ covered by COSMOS. This issue is one of
several described above that can in principle be addressed with
improvements to the simulations.

\begin{figure*}
	\centering
	\includegraphics[width=0.6\textwidth]{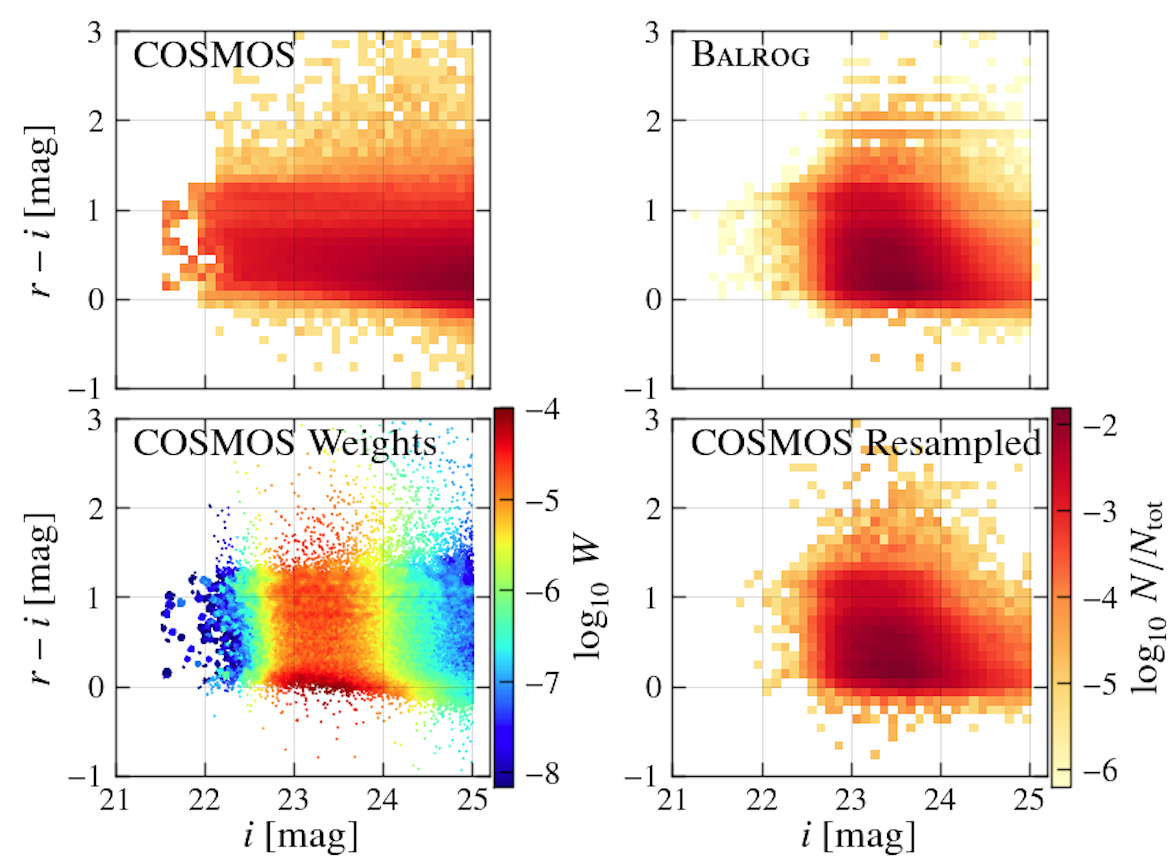}
	\caption{ COSMOS sample selection. The heat map colored
          histograms plot normalized counts.  {\bf Top left:} $i$-band
          magnitudes and $r-i$ colors for the full COSMOS catalog
          after basic quality cuts. {\bf Top right:} Distribution of
          $i$-band magnitude and $r-i$ colors using {\it truth catalog
            properties} of \balrog{} galaxies in our faint
          sample. {\bf Bottom left:} (Unnormalized) weights applied in
          the $i$, $r-i$ color plane to COSMOS galaxies in order to
          match the DES truth distribution. {\bf Bottom right:}
          $i$-band magnitudes, $r-i$ colors of the reweighted COSMOS
          sample. }
	\label{fig:cselect}
\end{figure*}

There are more fundamental limitations to this procedure,
however. When a simulated galaxy and a real galaxy overlap, it is not
always possible to determine whether the resulting catalog entry
belongs in the \balrog{} catalog. If the real object is largely
unmodified by the presence of the simulated galaxy, then associating
it with the truth properties of the simulated galaxy results in an
incorrect measurement of $\mathcal{L}$. If the real object is
substantially modified by the presence of the simulated galaxy, the
resulting catalog entry could be used to infer the
likelihood function for blends, though we have not built the inference
machinery necessary to do so. Finally, if the simulated object's
properties are not substantially modified by the presence of the real
object, then associating the resulting catalog  entry to the simulated
object's truth properties results in a useful measurement of
$\mathcal{L}$ at that location.

These ambiguous matches tend to introduce a small amount of real
galaxy contamination into the randoms, and result in a small
multiplicative bias to the clustering of roughly twice the
contamination rate. Excluding them excludes some \balrog{} galaxies in
a manner that reverses the sign of the multiplicative bias, with
similar amplitude. Ambiguous matches comprise only $\sim$$\, 1\%$ of our
\balrog{} galaxies, resulting in a multiplicative bias that is smaller
than the statistical error on the amplitude of the $w(\theta)$
measurement presented below. For this reason, we do not apply any
correction for this effect.

Finally, and most fundamentally, \balrog{} samples the likelihood
under slightly different conditions than the real data. If the image
contains $n$ real objects, the measurement likelihood for the $n^{th}$ is
\begin{align}
\mathcal{L} = p(\boldsymbol{\alpha}_{n,\rm meas}\: | \: \boldsymbol{\alpha}_{t,1},\boldsymbol{\alpha}_{t,2},...,\boldsymbol{\alpha}_{t,n-1}, \boldsymbol{\theta}),
\end{align} 
while the likelihood sampled in this image by a single added \balrog{}
galaxy is:
\begin{align}
\mathcal{L} = p(\boldsymbol{\alpha}_{n+1,\rm meas}\: | \: \boldsymbol{\alpha}_{t,1},\boldsymbol{\alpha}_{t,2},...,\boldsymbol{\alpha}_{t,n-1},\boldsymbol{\alpha}_{t,n}, \boldsymbol{\theta}).
\end{align} 
If the likelihood really is strongly non-local -- that is, if the
measured properties of each galaxy depend strongly on the properties
of other nearby objects -- then the \balrog{} catalogs will not be
sampling the same likelihood as the data, and we should not expect
$w(\theta)$ estimates made with them to be correct. All correlation
function estimators that use random catalogs assume that the window
function and the density field are statistically independent, however,
so a coupling between $\mathcal{L}$ and the galaxy density field would
also make \autoref{eq:ls} invalid for {\it any} random catalog.

These complications should all be much less severe for catalogs made
with the high-resolution space-based COSMOS imaging. Insofar as this
is true, we can regard any measured difference between the COSMOS
angular clustering and that measured with \balrog{} as evidence that
the simulated catalogs are not sampling the same likelihood function
as the data.

\subsection{Estimation algorithms}
\label{sec:est-alg}

We adopt the \citet{landy-szalay} estimator for the correlation function:
\begin{equation}
w\at{\theta} = \frac{DD - 2DR + RR}{RR},
\label{eq:ls}
\end{equation}
with $D$ labeling the data and $R$ labeling the randoms.  The
randoms sample the window function for an intrinsically unclustered
sample, and are used to remove any signal induced by non-uniform
detection probability. For our DES data, we will compare estimates of $w(\theta)$ made
using \balrog{} randoms to the same measurements using uniform randoms
that sample the survey geometry only 
(by applying the same spatial masking to the uniform randoms as applied to the data).
We have not run \balrog{} on the COSMOS imaging, and hence all
our COSMOS $w(\theta$) measurements use the standard uniform randoms.

We compute \autoref{eq:ls} using \textsc{TreeCorr}
\citep{treecorr}, a software package implementing a $k$-d tree
algorithm for efficient calculation of correlation functions over
large datasets.  We adjust the \texttt{bin\_slop} parameter, which
controls the fraction of the bin width by which pairs are allowed to
miss the correct bin, such that $\texttt{bin\_slop} \times
\texttt{bin\_size} < 0.1$, in order to reduce the binning errors made
by the algorithm.  We run \textsc{TreeCorr} over each of the 24 \km{}
\jk{} realizations, as explained in \autoref{sec:jk}, in order to
estimate the correlation function's covariance.

As a cross-check, we have also computed our correlation functions with
\textsc{athena} \citep{athena}, another tree-code which implements its
own internal \jk{} algorithm to estimate the covariance, where the
data's area is divided into squares on a grid of $N$~rows $\times$
$M$~columns, leaving out one of the squares in each \jk{} iteration.
Using either code, we measure consistent $w(\theta)$ signals.

As discussed in \citet{benchmark}, \jk{} resampling is a noisy
estimate of the covariance of $w(\theta)$, which is reasonably
well-suited for the diagonal elements, but theory-based errors are
better-suited for the off-diagonal terms.  Because we attempt no
physical interpretation of our clustering signals, we omit any
theoretical modeling, and do not explore noise estimates beyond \jk{}
resampling.

\subsection{DES sample selection}
\label{sec:DESsample}

We choose two separate DES samples for our clustering measurements: a
{ bright} sample ($21 < \texttt{MAG\_AUTO\_I} < 22$), which is a subset of the
magnitude selection used in the DES benchmark clustering analysis of
\citet{benchmark}, and a faint sample ($23 < \texttt{MAG\_AUTO\_I}
< 24$), where the DES catalogs are substantially incomplete, and, as we
will see in \autoref{sec:wmeas}, the variation in the observed galaxy
density across the sky is dominated by variations in the selection
function.  We should expect the {bright} clustering signal measured
with \balrog{} randoms to easily reproduce the signal measured with
uniform randoms (as done in the DES benchmark clustering
analysis) and to agree with COSMOS; this is primarily a sanity check.
Our faint selection is a strong test of the methodology -- success
here would indicate accurate measurement of spatial clustering even
where, because of the low signal-to-noise ratio of the sample,
anisotropies in the window function strongly affect the intrinsic
clustering signal.
Neither sample is identical to the DES benchmark sample;
in \autoref{sec:bscomp} we offer a brief look at this sample.

\subsection{COSMOS sample selection}
\label{sec:csample}

We use the public COSMOS multi-wavelength photometry catalog
\citep{capak2007} to validate our clustering measurements.  First, we
make a few basic quality cuts, selecting objects with:
\begin{align*}
&\texttt{blend\_mask = 0}  \\
\texttt{AND} \;\; &\texttt{star = 0}  \\
\texttt{AND} \;\; &\texttt{auto\_flag > -1}.
\end{align*}
At the time of this writing, we did not have an appropriate angular
mask for the COSMOS field. We have used the positions of objects
flagged as problematic in the COSMOS photometric catalog as our mask
definition. When constructing our sample, we first exclude any COSMOS
galaxy within $10\arcsec{}$ of an object flagged as bad. Visual
inspection shows good agreement between this set of bad objects and
problematic regions in the COSMOS imaging. Unfortunately, this
shortcut makes the small-scale COSMOS clustering difficult to
interpret, so we elect not to use COSMOS measurements of $w(\theta)$
for $\theta < 10\arcsec{}$ in the analyses.  We have increased the
10\arcsec{} separation cut, and verified that our results on scales
larger the masking radius are not sensitive to the value chosen.

Small changes in the properties of the selected galaxies can have
significant effect on the amplitude of $w(\theta)$, so we take care to
ensure that the sample we select from COSMOS is well-matched to the
DES galaxies.  Our technique for doing this is a resampling scheme
based on and motivated by that described in e.g. \citet{lima2008,
  sanchez2014}, and analogous to how we reweighted our \sersic{}
catalog in \autoref{sec:inputcat}.

First, we make the same cuts on the \balrog{} galaxies as we have for
the DES galaxies (cf. \autoref{sec:DESsample}).  For each
\balrog{} galaxy, we also have the truth magnitudes and colors used to
generate the galaxy, which are directly comparable to the magnitudes
and colors from the COSMOS photometric catalog
(cf. \autoref{sec:inputcat}).  Matching the properties of the
\balrog{} and COSMOS catalogs in this space should ensure similar
samples with comparable clustering.  We choose to work in
two dimensions: $i$-band magnitude and $r-i$ color, selecting
\texttt{i\_mag\_auto} and ($\texttt{r\_mag} - \texttt{i\_mag}$)%
\footnote{\texttt{i\_mag\_auto} quantifies a total magnitude, while
  $\texttt{r\_mag}$ and $\texttt{i\_mag}$ are 3\arcsec{} aperture
  measurements.}  from the COSMOS catalog as the complements to our
\balrog{} truth quantities.  The top row of \autoref{fig:cselect}
presents the COSMOS measurements alongside our faint \balrog{}
selection for the chosen quantities.

To match the samples, for each COSMOS galaxy we calculate the distance
to the $50^{\rm th}$-nearest \balrog{} galaxy in this color-magnitude
space.  The number of COSMOS galaxies inside this distance is
proportional to the ratio of the two distributions, and when properly
normalized, equal to the weight required to match them.  Normalization
is such that the ensemble of weights sums to unity.  We then randomly
resample the COSMOS catalog, using the calculated weights as the
selection probability for each object,%
\footnote{We resample to five times the number of objects with nonzero
  weights. However, results are insensitive to this choice; upping the
  sampling density arbitrarily high is unnecessary.}  which generates
our DES-matched COSMOS sample.

We repeat this process separately for both the bright sample and the
faint sample; \autoref{fig:cselect} presents our results for the faint sample.  Using the
weights in the bottom left panel, we resample the COSMOS catalog in
the top left panel.
After doing so, we recover the bottom right panel, which is
a good match to the top right panel -- the faint \balrog{} sample.  
We have confirmed that, after
this matching, the $g-$ and $z-$band magnitude distributions are also
strikingly similar to the \balrog{} truth distributions.  We have also
matched on quantities other than $r - i$ color and $i$-band magnitude,
as well as varied the number of nearest neighbors to query, and
measured consistent clustering signals.

\subsection{Stellar contamination}
\label{sec:contam}

\begin{figure}
	\centering
	\includegraphics[width=0.40\textwidth]{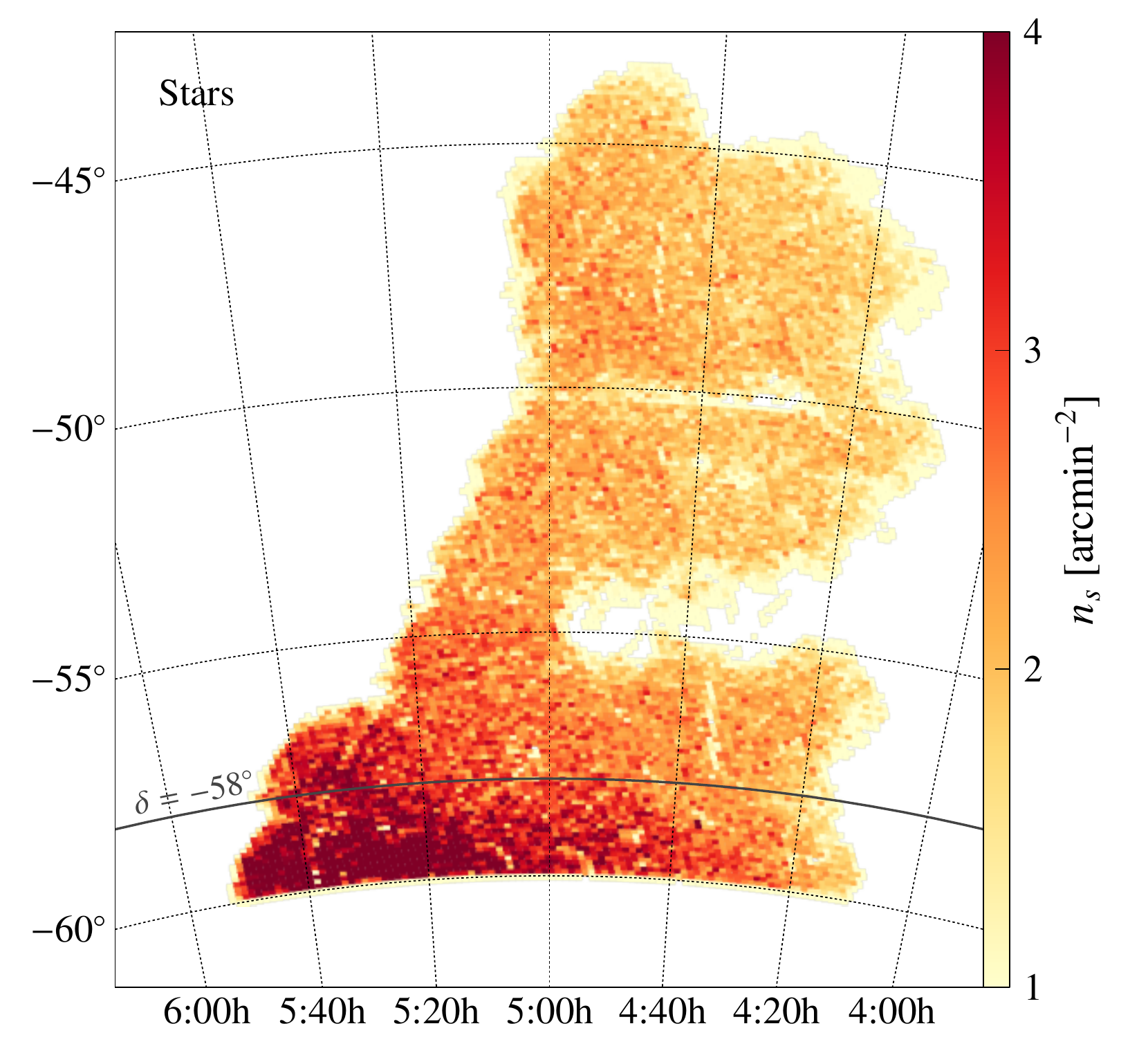}
	\caption{Map (declination vs. right ascension) of the DES
          stellar number density across the \spte{} footprint.  An
          additional parallel has been drawn at $\delta = -58\degr{}$,
          indicating the cut we make in our clustering measurements to
          eliminate the area of highest stellar contamination.  }
	\label{fig:starmap}
\end{figure}

Stars that are accidentally included in the galaxy clustering analysis
can have a significant impact on the measured clustering
\citep[e.g.,][]{SDSS_EDR, old_school}.  An unclustered stellar population
simply dilutes the measured angular clustering. If the stars
themselves cluster nontrivially, the measured signal is a mixture of
the true galaxy and stellar clustering, with mixture coefficients set
by the fraction $f$ of the galaxy sample that has been mis-classified
as stars. We refer readers to Appendix D of \citet{benchmark} for a 
detailed treatment of the subject.

To estimate the stellar contamination in our DES samples, we use the
\balrog{} simulations.  From the \balrog{} truth catalog, we can infer
the fraction of \balrog{} objects which were simulated as stars but
misclassified as galaxies.  However, because the DES and \balrog{}
stellar densities vary (cf. \autoref{sec:1d}), we need to renormalize
this \balrog{} contamination rate; we multiply by the detected DES
star-to-galaxy number ratio and divide by the detected \balrog{}
star-to-galaxy number ratio.

\begin{figure*}
	\centering
	\includegraphics[width=0.95\textwidth]{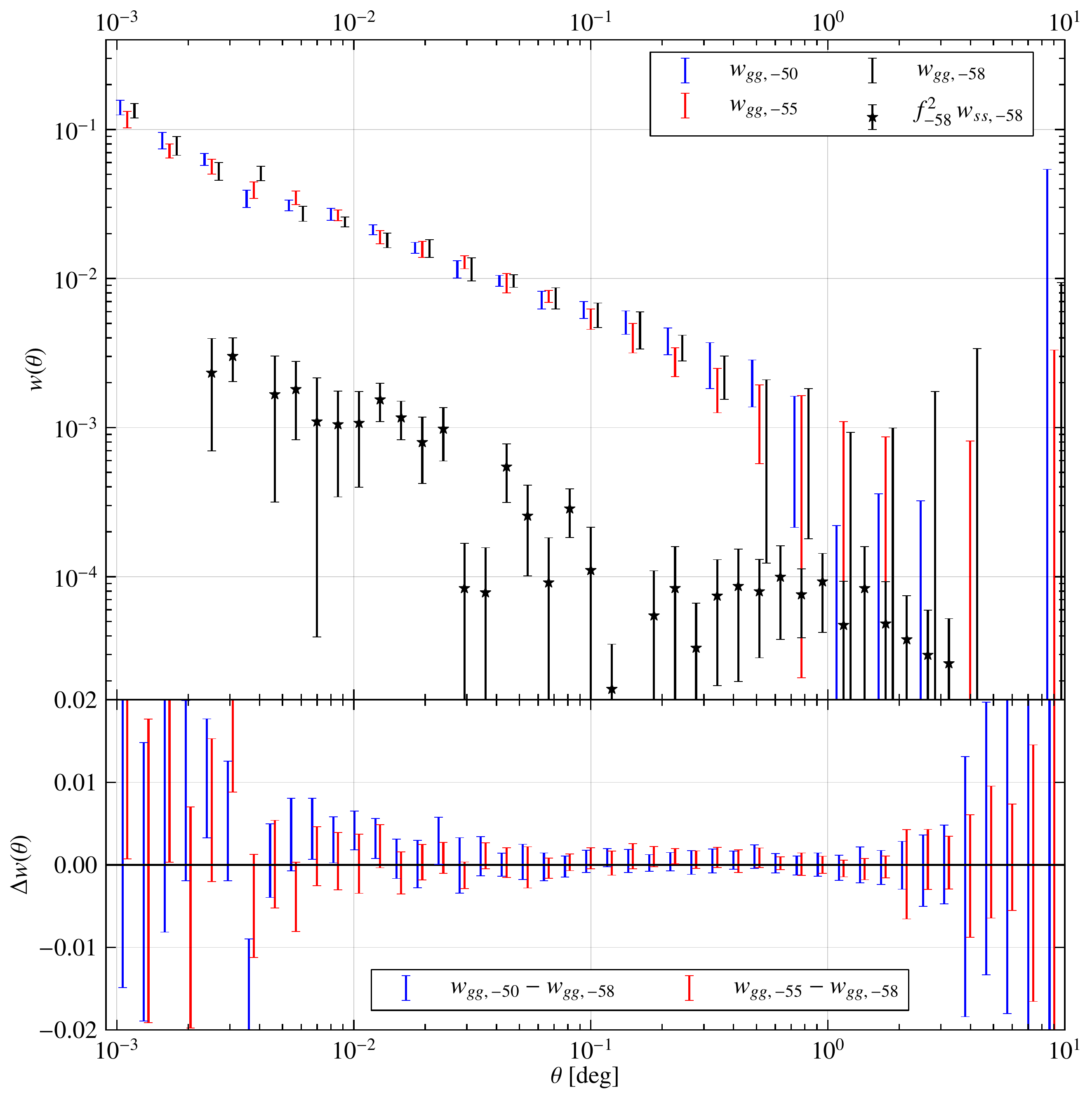}
	\caption{Testing stellar contamination.  All error bars in the
          figure are estimated with jackknife resampling
          (cf. \autoref{sec:jk}).  \textbf{Top:} The bar-only points
          show galaxy angular correlation function measurements for
          our faint ($ 23 < \texttt{MAG\_AUTO\_I} < 24$) DES sample
          over different declination ranges: $\delta > -50$ in blue,
          $-55 < \delta < -50$ in red, and $-58 < \delta < -55$ in
          black.  (For visual clarity, only every other point has been
          plotted, and there is a slight offset between points at the
          same angular scale. Legend labels denote the southern edge
          of the regions.)  Stellar density varies between the regions
          (cf. \autoref{fig:starmap}), and a stellar contamination
          dilution correction has been applied to each curve
          (cf. \autoref{sec:contam}).  The contamination fractions for
          each region are: $f_{-50} = 0.044$, $f_{-55} = 0.048$,
          $f_{-58} = 0.058$.  The black stars plot the stellar
          autocorrelation function multiplied by the square of the
          galaxy stellar contamination fraction, in the region of
          highest stellar density and highest stellar clustering.  (To
          maintain readability, we omit the stellar autocorrelations
          over the other two regions, and choose to focus on the most
          pessimistic case.)  If large enough, the stellar
          autocorrelation quantity can induce an additive bias to the
          galaxy clustering measurements, and we note that it is
          comparably small over the range of scales where we are able
          to make a statistically significant measurement.
          \textbf{Bottom:} Differences between the stellar
          contamination dilution corrected galaxy autocorrelation
          function measurements in the top panel.  There is no
          significant difference between the resulting measurements,
          suggesting that stellar contamination is not a significant
          source of systematic bias for this measurement.  }
	\label{fig:regions}
\end{figure*}

In both the bright and faint DES samples, we find $f \sim\!5\%$.
Inspection of the magnitude-FHWM plane in the COSMOS data indicates
that stellar contamination is small ($\sim\!0.1\%$ for $i < 22$), so
we omit any corrections due to this contamination in the COSMOS
measurements.

As shown in \autoref{fig:starmap}, the stellar density varies
dramatically across the DES survey area examined in this analysis. The
edge of the LMC intrudes at $\delta < -58$, so we have removed this
extreme region from the clustering analysis, and for the following
tests we divide the remainder of the area into three
declination-selected strips:
\begin{enumerate}
\item $\delta  > -50$,
\item $-55 < \delta < -50 $,
\item $-58 < \delta < -55 $,
\end{enumerate}
\noindent in order to test if our clustering signals are robust against stellar population size.
The two northernmost regions are roughly equal in stellar density,
while the southernmost's is about 35\% greater.

We measure the stellar autocorrelation $w_{ss}$ in each of the
declination-selected samples. The expected spurious clustering from
stellar contamination is proportional to this signal, but suppressed
by the square of the contamination fraction \citep{myers2006,
  benchmark}.  We find that $f^2 w_{ss}$ is well below errors in the
angular correlation function for both the bright and faint samples;
the faint measurements, which have larger stellar clustering, as well
as slightly higher stellar contamination, are shown in
\autoref{fig:regions}.%
\footnote{\texttt{MODEST\_CLASS} stellar selection is not entirely
  pure at $23 < \texttt{MAG\_AUTO\_I} < 24$, so a portion of the
  plotted stellar signal is actually from galaxies. We have also
  selected brighter magnitude ranges where the stellar selection is
  pure and found $f^2 w_{ss}$ to be smaller than what is shown in
  \autoref{fig:regions}; i.e. we have plotted the most pessimistic
  signal.  At any rate, even if our plotted $f^2 w_{ss}$ were more
  than a factor of 2 underestimated, it would still be below the level
  of errors in the galaxy-galaxy autocorrelation functions.}  (For
visual clarity, \autoref{fig:regions} only plots $f^2 w_{ss}$ in the
southernmost region, the most pessimistic case).  To account for
dilution from stellar contamination, we apply a $(1 + f)^2$ correction
\citep{myers2006, benchmark} to the galaxy autocorrelation functions.
We show in the bottom of \autoref{fig:regions} that after applying the
correction, the differences between the galaxy signals for the three
regions are small compared to the autocorrelation errors, further
indicating that stellar contamination is not a significant source of
systematic bias.

\subsection{Clustering measurements}
\label{sec:wmeas}

\begin{figure*}
	\centering
	\includegraphics[width=0.92\textwidth]{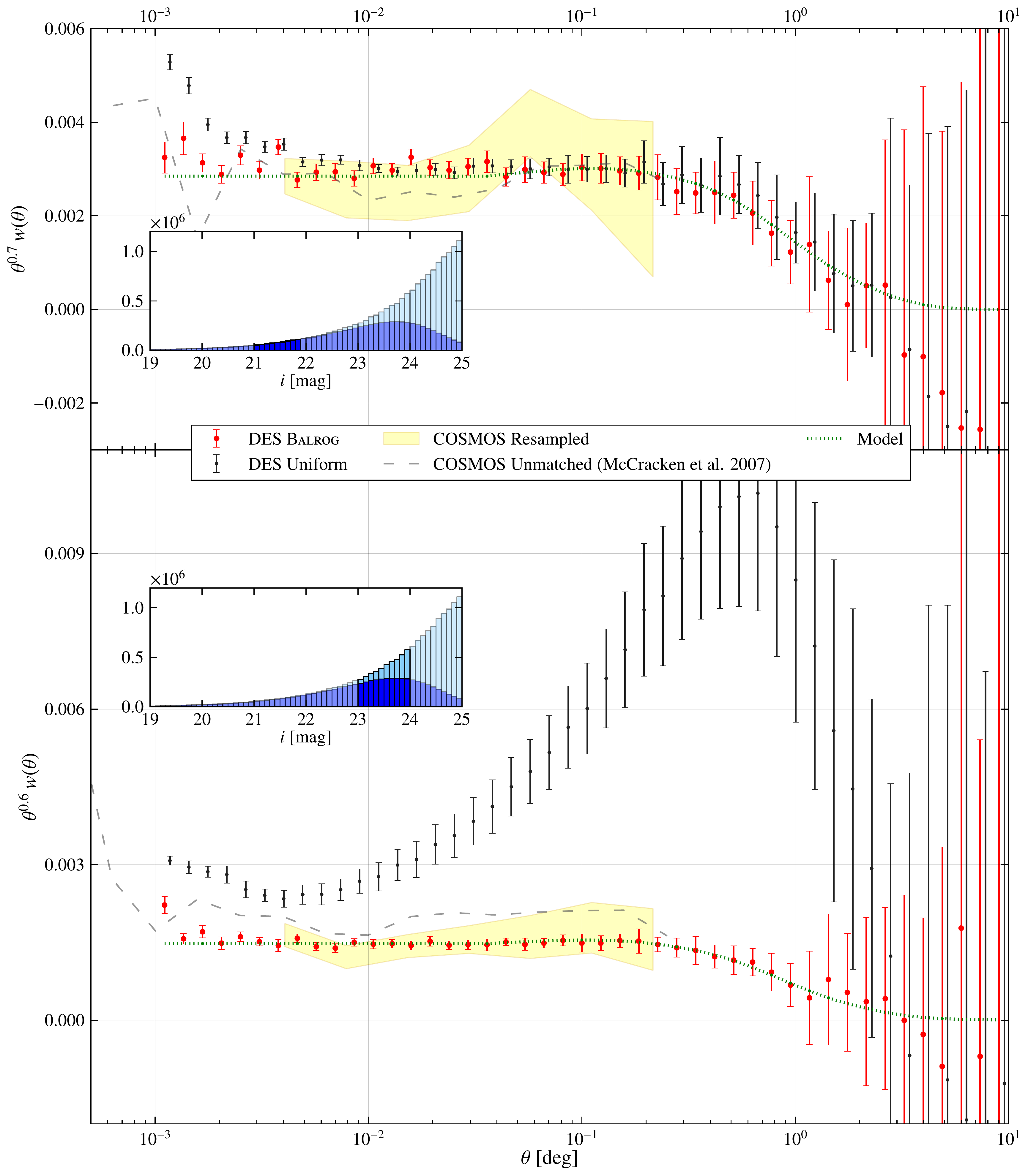}
	\caption{Angular clustering results. Black and red points show $w(\theta)$ measurements
          for our DES galaxies, with uniform and Balrog randoms,
          respectively.  (Points at the same separation have been
          slightly offset for visual clarity.)  The yellow band
          measures the 1$\sigma{}$ confidence interval on
          $w\at{\theta}$ in a matched COSMOS sample
          (cf. \autoref{sec:csample}).  All errors are
          estimated with \jk{} resampling, (see \autoref{sec:jk}).
	  The gray dashed lines are
          COSMOS measurements from \citet{COSMOS_McCracken}, which we
          note are \textit{not} matched to the DES sample, but which
          could be measured to a smaller scale than our DES-matched
          COSMOS measurements.  (See \autoref{sec:csample} and
          \autoref{sec:wmeas} for more details).  
	  Dashed green lines are $\Lambda$CDM model predictions,
	  not fits to the data (cf. \autoref{sec:wmeas}).
          Insets show the distribution of true \balrog{} (light blue)
          and observed DES (blue) magnitudes, with selection regions
          highlighted.  In both panels, we have multiplied the signal
          by its approximate power-law slope.  {\bf Top:} Clustering
          of the bright, fairly complete sample.  As expected,
          variations in the DES window function, as measured by the
          \balrog{} randoms, do not appear significant for the
          clustering above $15$\arcsec{} ($0.004 \degr$).  {\bf
            Bottom:} Clustering of the faint sample, which is near or
          at the magnitude limit of the survey, and $\sim\!35\%$
          incomplete on average. It is strongly impacted by
          systematic effects due to the spatial variations of DES survey
          properties.  We include the measurement using uniform
          randoms purely as an estimate of the of the importance of
          systematic errors, noting that it would be inappropriate to
          use uniform randoms to measure $w\at{\theta}$ for a $23 < i
          < 24$ sample selected with 10$\sigma$ limiting magnitude $i
          > 22.5$.  The \balrog{} randoms appear to capture
          essentially all of the extra power, suppressing it by
          roughly two orders of magnitude (see \autoref{sec:wmeas} for further explanation).
          Note the excellent agreement with the matched COSMOS
          measurements. Like \citet{COSMOS_McCracken}, \balrog{}
          suggests little deviation from a power-law down to small
          scales. The shape of \balrog{} results also agree with the shapes of the models.}
	\label{fig:clus}
\end{figure*}

\begin{figure*}
	\centering
	\includegraphics[width=0.65\textwidth]{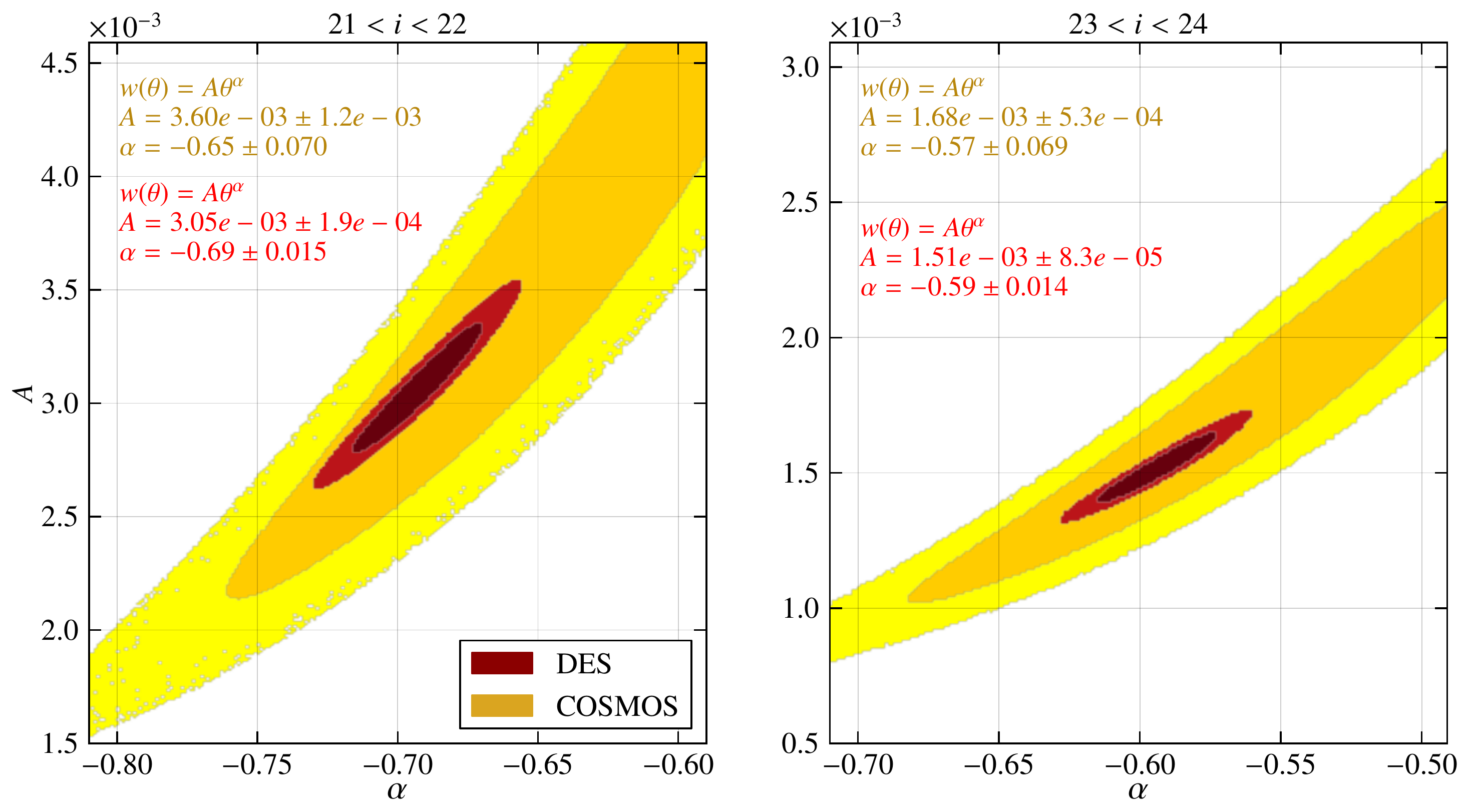}
	\caption{MCMC power-law fits for the $w(\theta)$ measurements shown in \autoref{fig:clus}.
Contours are the 68\% and 95\% intervals.
The DES measurements (red) use \balrog{} randoms, and the COSOMS measurements (yellow) are for the sample matched to DES.
The text displays the best-fit marginalized parameter values.
\textbf{Left:} bright sample. \textbf{Right:} faint sample.}
	\label{fig:fit}
\end{figure*}

We now present our $w\at{\theta}$ measurements.  Angular clustering
measurements for flux-limited samples generally see power-law behavior
at small angular separations, steepening above degree scales
\citep[e.g.][]{old_school, SDSS_EDR, COSMOS_McCracken}. We expect that
significant residual additive systematic errors should produce a deviation
from a constant power-law behavior below degree scales, while residual
multiplicative biases should produce a corresponding multiplicative
offset between the DES and COSMOS measurements.

Our bright sample galaxies are a subset of the DES benchmark sample,
which has been extensively studied in a separate analysis
\citep{benchmark}. The limiting magnitude of the benchmark sample
($i<22.5$) was made, in the conservative tradition of large-scale
structure measurements, in order to produce a clean sample with
relatively uniform selection; as shown in \autoref{fig:clus},
this selection indeed produces a reliable clustering signal at
large scales.

The top panel of \autoref{fig:clus} shows measurements of the angular
clustering for our bright ($21 < i < 22$)
sample. We plot $w\at{\theta}$ estimated using \balrog{} randoms in
red, and that estimated using the uniform randoms in black.  An overall correction to
the amplitude of both these DES curves has
been applied in order to correct for the effects of stellar dilution (cf. \autoref{sec:contam}).
The shaded
region shows the $1\sigma$ confidence interval (inferred from \jk{} resampling, cf. \autoref{sec:jk})
from the matched COSMOS photometric sample.

These three estimates are statistically consistent with one another
within the range probed by our COSMOS clustering measurement. Any
excess systematic power traced by the \balrog{} catalogs here is
evidently not significant for the measurements above $\theta \gtrsim
15\arcsec$ ($0.004 \degr$). Below this scale, the uniform and
\balrog{} curves diverge; the measurements made using the \balrog{}
sample continue the power-law behavior down to $\sim\!7''$, where
blending effects start to become significant. We have not attempted to
diagnose this behavior in detail.  However, we remark that COSMOS
measurements made by \citet{COSMOS_McCracken} for a similar, but not
identical sample, also suggest little deviation from a power-law down
to these scales; we include their measurements with our results in
\autoref{fig:clus}.  They select the same range of $i$-band
magnitudes, but we note that the sample is \textit{not} reweighted to
match the DES one (cf. \autoref{sec:csample}), and thus need not
exhibit an identical signal.  Therefore, the \citet{COSMOS_McCracken}
results offer strong evidence, but not definitive proof, to validate
the small-scale power-law-like \balrog{} results.

Our faint sample ($23 < i < 24$) is close to the
formal limiting magnitude for the survey. As is evident from
\autoref{fig:mags}, DES is substantially incomplete in this regime,
and this is where we should expect the spatial variation in survey
properties to matter the most. We include the clustering signal
measured using uniform randoms purely as an estimate of the of the
importance of systematic errors for this faint sample.

The bottom panel of \autoref{fig:clus} presents our angular clustering
results for this faint selection.  \balrog{} and the faint-sample
matched COSMOS results are in excellent agreement, and the former
continues its power-law behavior down to almost $4\arcsec{}$
(0.001\degr{}).  Subject to the same caveats discussed above, we again
plot a COSMOS measurement from \citet{COSMOS_McCracken}, using an
unmatched sample over the same magnitude range, noting similar
power-law behavior down to small scales.

The amplitude of the signal in the faint clustering measurement
closely follows our COSMOS signal. We note that the systematic error
has a substantially different shape than the galaxy autocorrelation,
and so where it is significant, it should produce a deviation from the
power-law behavior. This suggests that the residual additive
systematic error in the faint sample \balrog{} measurement is small
compared to the latter's jackknife errors. At $0.5\degr$, the
\balrog{} clustering errors are $\sim\!0.0005$, and so the spurious
clustering has been suppressed by about two orders of magnitude from
its value ($\sim\!0.01$ at the peak of the gray curve in
\autoref{fig:clus}).

%
%
%

To show that the shape of our clustering measurements matches general expectations,
we have included model $w(\theta)$ curves -- the dotted green lines in \autoref{fig:clus} -- for $\Lambda$CDM cosmology ($\sigma_{8}=0.8$, $\Omega_{m}=0.31$). 
These have been generated assuming the broad $dN/dz$ used in \citet{nock2010, aross2011} for a DES-like selection of galaxies.
For separations $r > 10$~Mpc/h,
we use a linear-theory correlation function, $\xi(r)$, 
derived by Fourier transforming the CAMB \citep{camb} power spectrum,
with $\xi(r) \propto r^{-\gamma}$ for $r < 10$~Mpc/h.
Projection to angular separations follows Equations~9-13 in \citet{crocce2011}. 
$w(\theta)$ was scaled by an arbitrary factor,
to account for galaxy bias and the true underlying dN/dz  
(both of which are expected to have nearly constant proportional effects on the amplitude as a function of $\theta$), 
with the curve set to be a power-law at $\theta < 0.03\degr$. 
In \autoref{fig:clus}, the shapes of the measured $w(\theta)$ curves indeed trace those of the model predictions.
In follow-up work, we will assess the impact on cosmological parameter sensitivity using our new methodology.
Here, the uncertainties in $w(\theta)$ at large angular scales, where cosmological sensitivity is the greatest,
are too large for us to draw interesting conclusions on the topic.

\autoref{fig:fit} plots the results when we fit power-laws to our $w(\theta)$ measurements:
\begin{equation}
w(\theta) = A \theta^{\alpha}.
\end{equation}
The darker contours show the 68\% confidence intervals on the amplitude ($A$) and the power-law index ($\alpha$),
while the lighter contours show the 95\% confidence intervals for these quantities.
We also indicate the best-fit (marginalized) parameter values in the figure.
The COSMOS results are those of the DES-matched sample, and
the DES results are calculated using the \balrog{} randoms. The fits are made
using \texttt{emcee} 
\citep{emcee},
an affine-invariant Markov chain Monte Carlo (MCMC) sampler. 
We find the off-diagonal components of the \jk{} covariance estimates to be unstable in the fits (cf. \autoref{sec:est-alg}; \citealt{benchmark}),
so we have restricted the $\chi^2$ likelihood sampling to diagonal elements only.
The fits extend over the range of angular scales probed by the COSMOS measurements ($0.004\degr < \theta < 0.2\degr$).

In both the bright and faint samples, the DES results fall inside the 1$\sigma$ COSMOS contours.
Owing to the much increased survey area, the DES measurements shrink the uncertainty contours considerably,
by about a factor of 5 or more in both $\alpha$ and $A$.
When we fix the power-law index to the best-fit DES value, and fit for the scaling amplitude between the two samples,
we find this amplitude to be $1.04 \pm 0.11$ in the bright sample, and $1.00 \pm 0.09$ in the faint sample.

\section{Discussion}
\label{sec:discussion}

We have developed a Monte Carlo injection simulation software package
designed to allow accurate inference of galaxy ensemble properties
where the catalogs are likely to be highly biased and incomplete. Our
simulations are computationally tractable, requiring approximately
$3~{\rm CPU}$~${\rm seconds}$ per simulated galaxy, and the resulting
catalogs have the same pattern of systematic variation with image
quality as the real data.

We demonstrate that the use of these simulated catalogs as randoms in
a clustering measurement is an effective and operationally simple way
to suppress systematic errors in the angular clustering signal. We use
\balrog{} catalogs generated with DES data to reproduce the known
angular clustering of faint galaxies previously measured with high
quality space-based imaging data. We show that this measurement agrees
with the COSMOS measurement, even for galaxies for which DES is
substantially incomplete.

\begin{figure}
	\centering
	\includegraphics[width=0.45\textwidth]{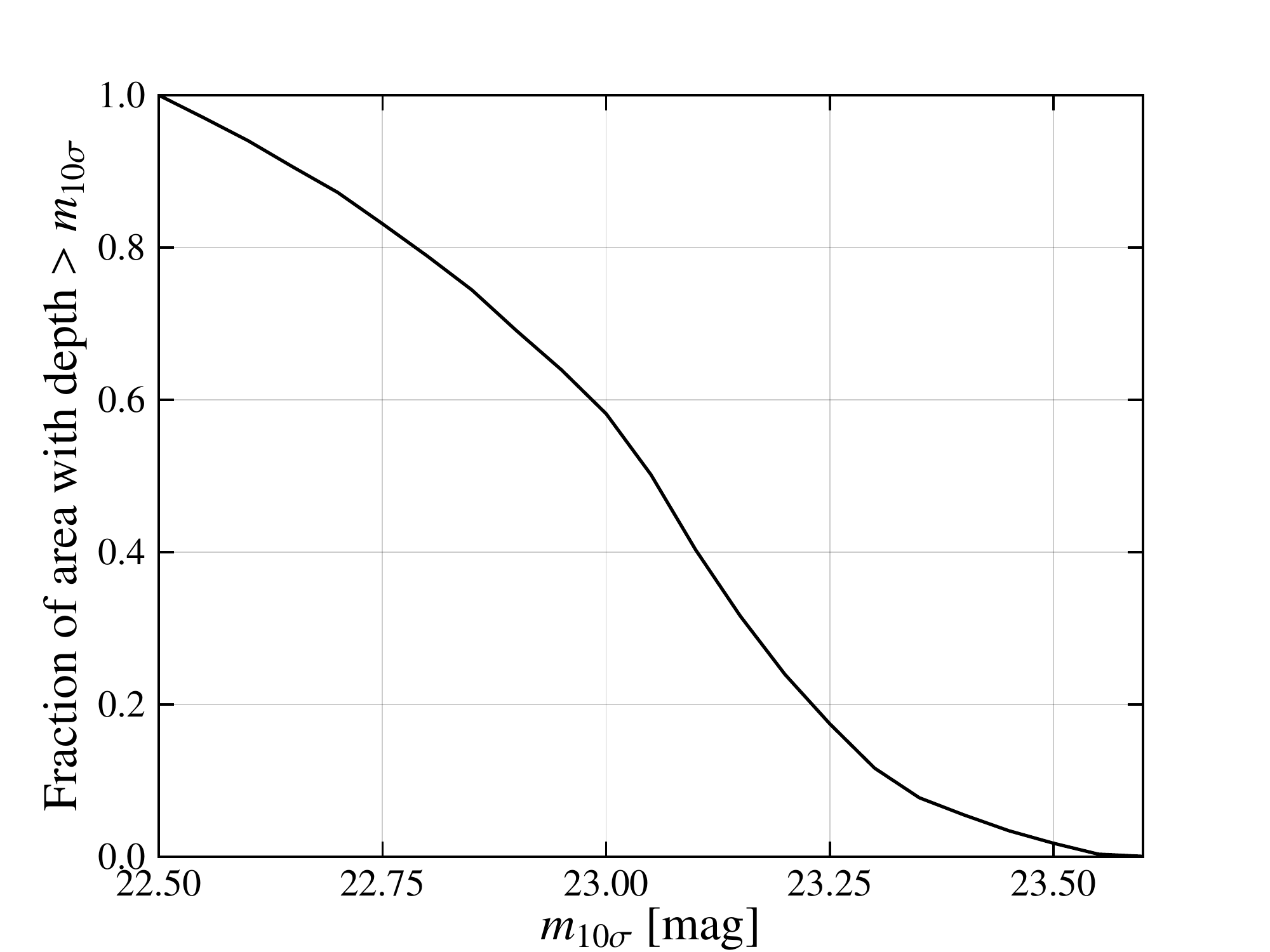}
	\caption{Area as a function of ($10\sigma$ $i$-band) depth for our DES
          clustering samples.  Traditionally clustering analyses
          select magnitudes $\leq$ to the depth.  We have included
          $\texttt{MAG\_AUTO\_I} < 24$ galaxies, beyond the 
          limiting magnitude of any or our area.}
	\label{fig:area}
\end{figure}

\autoref{fig:area} plots the area coverage of our DES sample as
function of depth.  In the conservative approach, clustering analyses
often select only galaxies brighter than the magnitude limit.  We have
included galaxies as faint as $\texttt{MAG\_AUTO\_I}=24$, for which
there is no area in our sample reaching this depth.

This procedure extends the reach of clustering measurements in
ground-based surveys like DES to much deeper samples, enabling
statistical science for rare, faint, and high-redshift objects near
the survey limit, fully exploiting the great data volume of the
surveys.  This is the first time, as far as we are aware, that
accurate angular clustering measurements have been made with a
substantially incomplete sample.

The data represented here are a small fraction of the final DES data
volume. In future work, we will generate \balrog{} catalogs covering
all the imaging data. Several simple improvements over the analysis
presented here are planned, including folding in photometric redshifts
into the measurements 
(see \citealt{bonnett2015, sanchez2014} as references describing photometric redshift estimation for DES); 
using an input catalog with galaxy colors
matched to the DECam filters; embedding the simulations into the full
stack of single-epoch images instead of directly into the coadds; and
adopting input catalogs spanning a larger range of galaxy properties,
in order to avoid the intrinsic sample variance of catalogs drawn from
the small COSMOS field.

We anticipate that injection simulations similar to \balrog{} will be
useful for a wide variety of measurements beyond clustering. Accurate
models of biases and completeness can, we hope, let modern surveys
take full advantage of all the available data.

\section*{Acknowledgements}


The authors are grateful to Chris Hirata and John Beacom for many
illuminating discussions, and to Todd Tomashek for guidance on
integrating the final catalogs into the Dark Energy Survey science database.
We commend the \galsim{} developers for their assistance and for exemplifying perhaps the best code documentation
throughout the astronomical community.
We thank An\u{z}e Slosar and the astrophysics group at Brookhaven National Laboratory
for use of computing resources throughout this work.
We are indebted to the entire DES Data Management team for the often under-appreciated hard work that they do.
We owe much gratitude to the late Steve Price for his beyond generous support of CCAPP for many years.

E.S. is supported by an Ohio State University Graduate Presidential Fellowship. 
E.M.H. is funded by a CCAPP postdoctoral fellowship. 
J.A. is partially supported by MINECO under grant FPA2012-39684.
P.M. is supported by the U.S. Department of Energy under Contract No. DE-FG02-91ER40690

We are grateful for the extraordinary contributions of our CTIO colleagues and the DECam Construction, Commissioning and Science Verification
teams in achieving the excellent instrument and telescope conditions that have made this work possible.  The success of this project also 
relies critically on the expertise and dedication of the DES Data Management group.

Funding for the DES Projects has been provided by the U.S. Department of Energy, the U.S. National Science Foundation, the Ministry of Science and Education of Spain, 
the Science and Technology Facilities Council of the United Kingdom, the Higher Education Funding Council for England, the National Center for Supercomputing 
Applications at the University of Illinois at Urbana-Champaign, the Kavli Institute of Cosmological Physics at the University of Chicago, 
the Center for Cosmology and Astro-Particle Physics at The Ohio State University,
the Mitchell Institute for Fundamental Physics and Astronomy at Texas A\&M University, Financiadora de Estudos e Projetos, 
Funda{\c c}{\~a}o Carlos Chagas Filho de Amparo {\`a} Pesquisa do Estado do Rio de Janeiro, Conselho Nacional de Desenvolvimento Cient{\'i}fico e Tecnol{\'o}gico and 
the Minist{\'e}rio da Ci{\^e}ncia, Tecnologia e Inova{\c c}{\~a}o, the Deutsche Forschungsgemeinschaft and the Collaborating Institutions in the Dark Energy Survey. 
The DES data management system is supported by the National Science Foundation under Grant Number AST-1138766.

The Collaborating Institutions are Argonne National Laboratory, the University of California at Santa Cruz, the University of Cambridge, Centro de Investigaciones En{\'e}rgeticas, 
Medioambientales y Tecnol{\'o}gicas-Madrid, the University of Chicago, University College London, the DES-Brazil Consortium, the University of Edinburgh, 
the Eidgen{\"o}ssische Technische Hochschule (ETH) Z{\"u}rich, 
Fermi National Accelerator Laboratory, the University of Illinois at Urbana-Champaign, the Institut de Ci{\`e}ncies de l'Espai (IEEC/CSIC), 
the Institut de F{\'i}sica d'Altes Energies, Lawrence Berkeley National Laboratory, the Ludwig-Maximilians Universit{\"a}t M{\"u}nchen and the associated Excellence Cluster Universe, 
the University of Michigan, the National Optical Astronomy Observatory, the University of Nottingham, The Ohio State University, the University of Pennsylvania, the University of Portsmouth, 
SLAC National Accelerator Laboratory, Stanford University, the University of Sussex, and Texas A\&M University.

The DES participants from Spanish institutions are partially supported by MINECO under grants AYA2012-39559, ESP2013-48274, FPA2013-47986, and Centro de Excelencia Severo Ochoa SEV-2012-0234.
Research leading to these results has received funding from the European Research Council under the European Union’s Seventh Framework Programme (FP7/2007-2013) including ERC grant agreements 
 240672, 291329, and 306478.

This paper has gone through internal review by the DES collaboration.
The document is identified as \texttt{FERMILAB-PUB-15-307-AE} and \texttt{DES-2015-0099}.


\bibliography{references}

\appendix
\section{Masking}
\label{sec:mask}

We apply the mask of \citet{benchmark} to our data.
This mask is made in a five-step process.

\begin{enumerate}

\item 
Coordinate cuts are made to select area in the SV \spte{} region (cf. \autoref{sec:desdata}).
The relevant cut for the area over which we have run \balrog{} is $\delta > -60$.
This avoids areas of high stellar density from the LMC. \newline

\item As mentioned in \autoref{sec:cats}, \sex{} detections include a
  population with large offsets between windowed centroid measurements
  in different bands.  The SV footprint was pixelized at HEALPix
  resolution $\texttt{NSIDE} = 4096$, masking the 4\% of the pixels
  with the highest density of objects with:
\begin{align*}
 \texttt{FLUX\_}&\texttt{AUTO\_G} / \texttt{FLUXERR\_AUTO\_G} > 5 \; \;   \texttt{AND} \; \; \\ 
&\lVert \, \, \, \, (\texttt{ALPHAWIN\_J2000\_G}, \texttt{DELTAWIN\_J2000\_G})  \\ &  -(\texttt{ALPHAWIN\_J2000\_I}, \texttt{DELTAWIN\_J2000\_I}) \, \, \, \, \rVert > 1\arcsec 
\end{align*}

About 25\% of the large outlier population is within these regions.  \newline

\item The mask eliminates areas in close proximity to bright stars
  from the 2MASS catalog \citep{twomass}.  A circular exclusion region is drawn around
  each 2MASS star with radius $(-10 \, M_{J} + 150)$\arcsec{}, where
  $M_{J}$ is the $J$-band magnitude, setting a maximum radius of
  120\arcsec{} and eliminating all circles with radius $<
  30$\arcsec{}.  The footprint is pixelized at $\texttt{NSIDE} = 4096$
  resolution, and HEALPixels whose centers fall within 10\arcsec{} of
  any exclusion zone are flagged as bad in the mask.  \newline

\item The mask selects regions with 10$\sigma$ limiting depth of
  $\texttt{MAG\_AUTO\_I} > 22.5$, where the depths are calculated
  according to procedure presented in \eliip{}.  Briefly, the \sex{}
  \texttt{MAGERR\_AUTO} vs. \texttt{MAG\_AUTO} distribution is fit in
  pixels of HEALPix resolution \texttt{NSIDE = 1024} to determine the
  depth on a coarse scale.  The random forest algorithm implemented in
  \textsc{sklearn}\footnote{\href{http://scikit-learn.org}{http://scikit-learn.org}}
  is used to find coefficients on this pixelation scale which fit the
  depth as a function of:
  \begin{enumerate}
\item the \texttt{MANGLE} \citep{mangle} 10$\sigma{}$ limiting magnitude measurements in 2\arcsec{} apertures available from DESDM,
\item maps of the survey observing properties (e.g. airmass, PSF size,
  etc.) compiled by \citet{boris_maps} (see also \autoref{sec:2d}).
\end{enumerate}
These products are generated at a finer resolution than the
\texttt{MAGERR\_AUTO} vs. \texttt{MAG\_AUTO} curve can be fit: the
maps of \citet{boris_maps} at \texttt{NSIDE = 4096}, and
\texttt{MANGLE} to arbitrary resolution, meaning the survey depth can then
be mapped more finely using the coefficients of these
quantities. \newline

\item The mask selects regions where at least 80\% of the area
  includes detections.  Each region is defined on a HEALPix grid of
  \texttt{NSIDE = 4096}, checking for detections within each of the
  64~subpixels of an \texttt{NSIDE = 32768} pixelized \texttt{MANGLE}
  mask.

\end{enumerate}

\section{Jackknife Errors}
\label{sec:jk}

\begin{figure}
	\centering
	\includegraphics[width=0.45\textwidth]{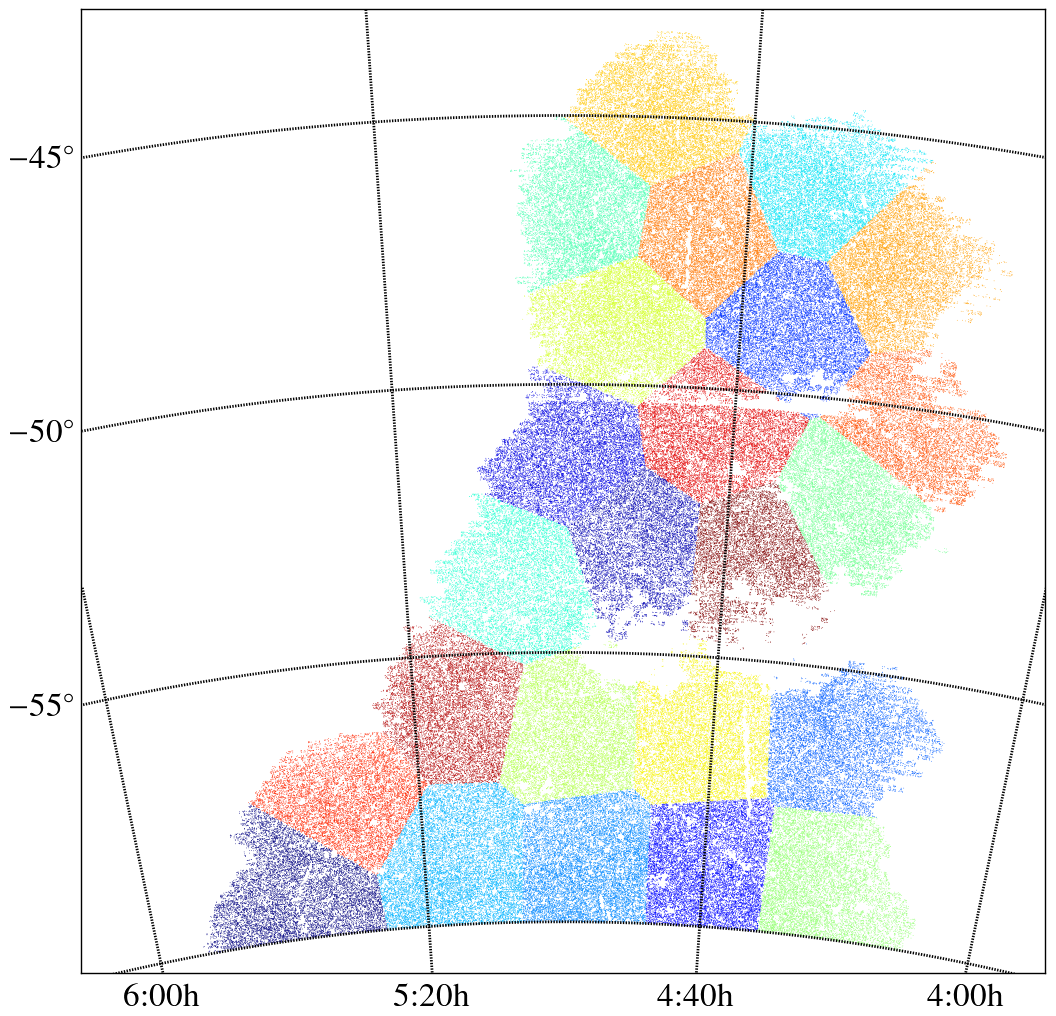}
	\caption{$k$-means \jk{} regions. Each point is a DES galaxy, 
colored according to which of the 24 $k$-means clusters it is assigned membership.
The algorithm divides the footprint into regions with roughly uniform cardinality.}
	\label{fig:km}
\end{figure}

Several instances of the work in this paper make use of \jk{} error
estimates.  We generate \jk{} regions for our data's footprint using a
\km{} algorithm,%
\footnote{\href{https://github.com/esheldon/kmeans\_radec/}{https://github.com/esheldon/kmeans\_radec/}}
a method to partition $n$~data points into $k$-clusters, assigning
each data point into the cluster with the nearest mean; here, the
region closest in angular distance.  The set of clusters,
$\boldsymbol{S}=\left\{ S_1, S_2, ..., S_k \right\}$, with centers
$\boldsymbol{\mu}=\left\{ \mu_1, \mu_2, ..., \mu_k \right\}$, is
generated by minimizing the within-cluster sum of distance squares:
\begin{equation}
  \underset{\boldsymbol{S}}{\mathrm{arg \, min}} \sum_{i=1}^{k} \sum_{\boldsymbol{x} \in S_i} \norm{\boldsymbol{x} - \boldsymbol{\mu}_{i}}^2.
\label{eq:km}
\end{equation}
Each datum is associated to the region whose center is nearest on the
celestial sphere, where a cluster's set of associated points has been
labeled as $\boldsymbol{x}$.  For approximately uniform data, \km{}
produces cluster sets roughly equal in number of associated points.
\autoref{fig:km} shows $k$-means classification for our DES galaxies,
after applying the cuts described in \autoref{sec:cats}; galaxies are
colored according to which cluster they were assigned.

After generating \km{} \jk{} regions, we proceed in the usual way to
estimate \jk{} errors.  One $S_n$ and its associated $\boldsymbol{x}$
is left out in each realization, and we find the covariance of the
vector of interest over the realizations:
\begin{equation}
  C_{ij} = \frac{(N-1)}{N} \sum_{n=1}^{N}  {\left[\vphantom{f_n\at{x_j} - f\at{x_j}}  f_n(x_i) - f(x_i)  \right] } {\left[ f_n(x_j) - f(x_j)  \right] },
\label{eq:jk-cov}
\end{equation}
where $f$ is the measurement over the full area, without removing any of the sample, 
and $f_n$ is the realization with $S_n$ removed.
$N$ is the number of \jk{} regions;
we use $N = 24$ throughout this work.

\section{Benchmark Comparison}
\label{sec:bscomp}

Some of the ongoing and planned clustering analyses of DES data make
use of the \textit{benchmark} sample, which is described in full in
\citet{benchmark}. This sample uses the mask described in
\autoref{sec:mask}. Galaxies are selected with a magnitude cut $18 <
\texttt{MAG\_AUTO\_I} < 22.5$. Star-galaxy separation is performed
using a new quantity, termed \texttt{WAVG\_SPREAD\_MODEL}, which is a
weighted average of the \sextractor{} \texttt{SPREAD\_MODEL} quantity
estimated from stars in the single-epoch DES images. \citet{benchmark}
measures the angular clustering of this sample, recovering results
that are in general agreement with prior measurements.

We present here an additional, approximate validation of the DES benchmark
results. Without \balrog{} galaxies embedded in single-epoch images, we
cannot perfectly capture the effects of the star-galaxy separation
used in selecting the benchmark sample. However, we measure and adjust
for the stellar contamination as in \autoref{sec:contam}, thus we do not
expect any substantial difference in the resulting measurement.

\begin{figure*}
	\centering
	\includegraphics[width=0.99\textwidth]{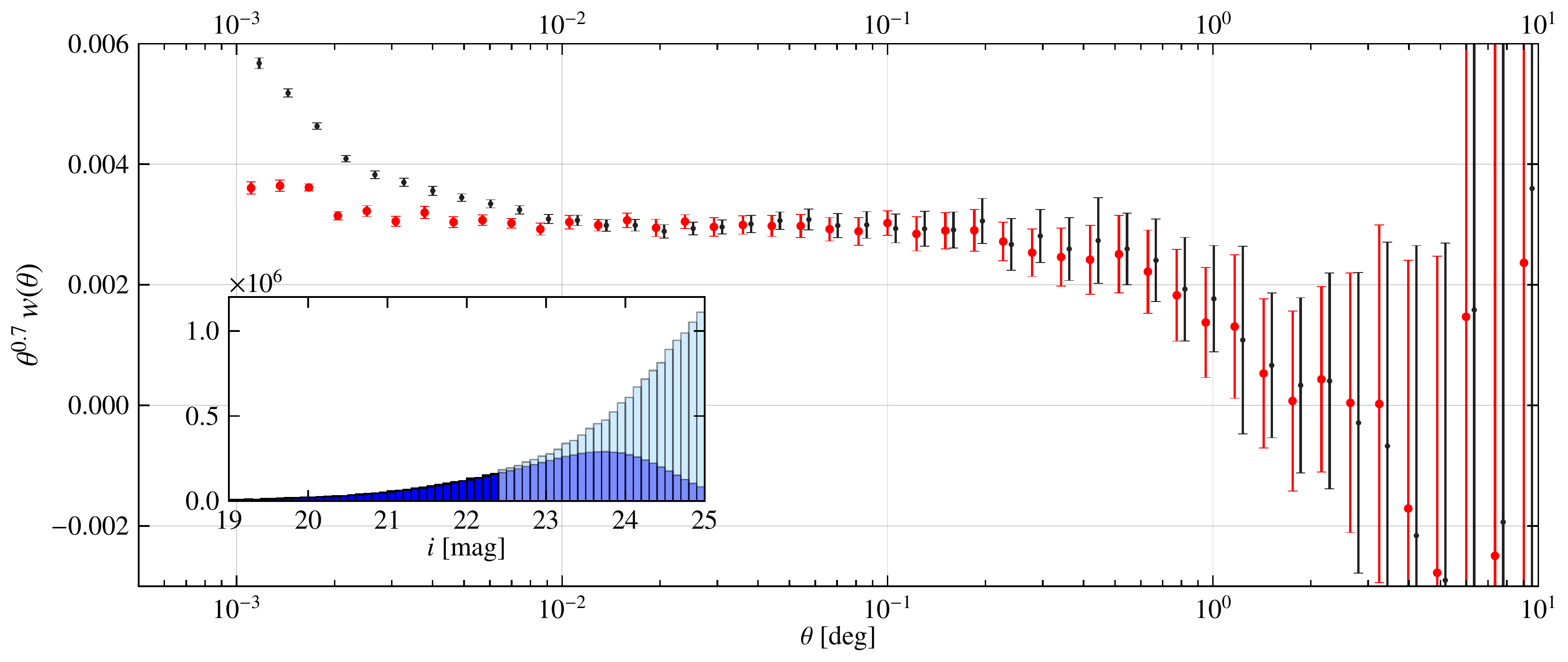}
	\caption{Angular clustering measurements using a sample
          similar to that of \citet{benchmark}, with \balrog{} (red
          points) and uniform randoms (black points). The figure is
          similar to \autoref{fig:clus}. Selection cuts are discussed
          in \autoref{sec:cats} and \autoref{sec:mask}. Shown in the
          inset, a magnitude cut of $18 < \texttt{MAG\_AUTO\_I} <
          22.5$ has been applied; blue is the observed magnitude
          distribution and light blue is the truth magnitude
          distribution from \balrog{}. The correlation functions have been scaled by
          the approximate power-law slope. The results suggest that
          the measurements made in \citet{benchmark} are unaffected by
          significant sources of systematic bias at scales $\theta >
          0.01\degr$.} 
	\label{fig:bmcomp}
\end{figure*}

A comparison between the clustering signals of our benchmark-like
sample, measured with uniform and with \balrog{} randoms, is shown in
\autoref{fig:bmcomp}. The results are quantitatively similar to those
shown in \autoref{fig:clus}. There is no significant correction
introduced by \balrog{} above $0.01\degr$, suggesting that the
benchmark sample is unaffected by significant measurement biases
at moderate and large scales. This is consistent with the independent
measurements from \citet{benchmark}.

\hypertarget{affil}{}
\section*{Author affiliations}
{\small
\begin{enumerate}[label=$^{\arabic*}\,$, leftmargin=*, align=left]
\item Department of Physics, The Ohio State University, Columbus, OH 43210, USA
\item Center for Cosmology and Astro-Particle Physics, The Ohio State University, Columbus, OH 43210, USA
\item Institut de F\'{\i}sica d'Altes Energies, Universitat Aut\`onoma de Barcelona, E-08193 Bellaterra, Barcelona, Spain
\item Department of Physics \& Astronomy, University College London, Gower Street, London, WC1E 6BT, UK
\item Jodrell Bank Center for Astrophysics, School of Physics and Astronomy, University of Manchester, Oxford Road, Manchester, M13 9PL, UK
\item Institut de Ci\`encies de l'Espai, IEEC-CSIC, Campus UAB, Carrer de Can Magrans, s/n,  08193 Bellaterra, Barcelona, Spain
\item Kavli Institute for Particle Astrophysics \& Cosmology, P. O. Box 2450, Stanford University, Stanford, CA 94305, USA
\item National Accelerator Laboratory, Menlo Park, CA 94025, USA
\item Brookhaven National Laboratory, Bldg 510, Upton, NY 11973, USA
\item Cerro Tololo Inter-American Observatory, National Optical Astronomy Observatory, Casilla 603, La Serena, Chile
\item Department of Physics and Electronics, Rhodes University, PO Box 94, Grahamstown, 6140, South Africa
\item Fermi National Accelerator Laboratory, P. O. Box 500, Batavia, IL 60510, USA
\item Institute of Astronomy, University of Cambridge, Madingley Road, Cambridge CB3 0HA, UK
\item Kavli Institute for Cosmology, University of Cambridge, Madingley Road, Cambridge CB3 0HA, UK
\item CNRS, UMR 7095, Institut d'Astrophysique de Paris, F-75014, Paris, France
\item Sorbonne Universit\'es, UPMC Univ Paris 06, UMR 7095, Institut d'Astrophysique de Paris, F-75014, Paris, France
\item Laborat\'orio Interinstitucional de e-Astronomia - LIneA, Rua Gal. Jos\'e Cristino 77, Rio de Janeiro, RJ - 20921-400, Brazil
\item Observat\'orio Nacional, Rua Gal. Jos\'e Cristino 77, Rio de Janeiro, RJ - 20921-400, Brazil
\item Department of Astronomy, University of Illinois, 1002 W. Green Street, Urbana, IL 61801, USA
\item National Center for Supercomputing Applications, 1205 West Clark St., Urbana, IL 61801, USA
\item Institute of Cosmology \& Gravitation, University of Portsmouth, Portsmouth, PO1 3FX, UK
\item George P. and Cynthia Woods Mitchell Institute for Fundamental Physics and Astronomy, and Department of Physics and Astronomy, Texas A\&M University, College Station, TX 77843,  USA
\item Excellence Cluster Universe, Boltzmannstr.\ 2, 85748 Garching, Germany
\item Faculty of Physics, Ludwig-Maximilians University, Scheinerstr. 1, 81679 Munich, Germany
\item Universit\"ats-Sternwarte, Fakult\"at f\"ur Physik, Ludwig-Maximilians Universit\"at M\"unchen, Scheinerstr. 1, 81679 M\"unchen, Germany
\item Department of Physics and Astronomy, University of Pennsylvania, Philadelphia, PA 19104, USA
\item Jet Propulsion Laboratory, California Institute of Technology, 4800 Oak Grove Dr., Pasadena, CA 91109, USA
\item Department of Astronomy, University of Michigan, Ann Arbor, MI 48109, USA
\item Department of Physics, University of Michigan, Ann Arbor, MI 48109, USA
\item Kavli Institute for Cosmological Physics, University of Chicago, Chicago, IL 60637, USA
\item Max Planck Institute for Extraterrestrial Physics, Giessenbachstrasse, 85748 Garching, Germany
\item Australian Astronomical Observatory, North Ryde, NSW 2113, Australia
\item Departamento de F\'{\i}sica Matem\'atica,  Instituto de F\'{\i}sica, Universidade de S\~ao Paulo,  CP 66318, CEP 05314-970, S\~ao Paulo, SP,  Brazil
\item Instituci\'o Catalana de Recerca i Estudis Avan\c{c}ats, E-08010 Barcelona, Spain
\item Centro de Investigaciones Energ\'eticas, Medioambientales y Tecnol\'ogicas (CIEMAT), Madrid, Spain
\item Department of Physics, University of Illinois, 1110 W. Green St., Urbana, IL 61801, USA
\item Argonne National Laboratory, 9700 South Cass Avenue, Lemont, IL 60439, USA
\item Department of Physics, Stanford University, 382 Via Pueblo Mall, Stanford, CA 94305, USA
\end{enumerate}
}

\end{document}